# Pitch-axis supermanoeuvrability in a biomimetic morphing-wing UAV

**Arion Pons[1,2,*] and Fehmi Cirak[2]**

[1] Division of Fluid Dynamics, Department of Mechanics and Maritime Sciences, Chalmers University of Technology, Gothenburg 412 96, Sweden
[2] Department of Engineering, University of Cambridge, Cambridge CB2 1PZ, UK

## Abstract

Birds and bats are extremely adept flyers: whether in hunting prey, or evading predators, post-stall manoeuvrability is a characteristic of vital importance. Their performance, in this regard, greatly exceeds that of uncrewed aerial vehicles (UAVs) of similar scale. Attempts to attain post-stall manoeuvrability, or supermanoeuvrability, in UAVs have typically focused on thrust-vectoring technology. Here we show that biomimetic wing morphing offers an additional pathway to classical supermanoeuvrability, as well as novel forms of bioinspired post-stall manoeuvrability. Using a state-of-the-art flight simulator, equipped with a multibody model of lifting surface motion and a delay differential equation (Goman-Khrabrov) dynamic stall model for all lifting surfaces, we demonstrate the capability of a biomimetic morphing-wing UAV for two post-stall manoeuvres: a classical rapid nose-pointing-and-shooting (RaNPAS) manoeuvre; and a wall landing manoeuvre inspired by biological ballistic transitions. We develop a guidance method for these manoeuvres, based on parametric variation of nonlinear longitudinal stability profiles, which allows efficient exploration of the space of post-stall manoeuvres in these types of UAVs; and yields insight into effective morphing kinematics to enable these manoeuvres. Our results demonstrate the capability of biomimetic morphing, and morphing control of nonlinear longitudinal stability, to enable advanced forms of transient supermanoeuvrability in UAVs.

[*] Corresponding author. *E-mail*: arion@chalmers.se

## Nomenclature

$t$ = time, s

$\theta$ = aircraft pitch, rad

$\Lambda$ = wing symmetric sweep, positive forward, rad

$\Gamma$ = wing symmetric dihedral, positive upward, rad

$\Phi$ = wing symmetric incidence angle, positive upward, rad

$\alpha$ = local effective angle of attack, rad

$\beta_e$ = elevator deflection, rad

$U$ = local airspeed, m/s

$c$ = local chord, m

$b$ = local semichord, m

$C_i$ = aerodynamic coefficient

$\tau_i$ = Goman-Khrabrov delay parameter

$p$ = Goman-Khrabrov mixing parameter

$p_0$ = Goman-Khrabrov quasistatic mixing function

$a_i, \ldots, e_i$ = aerodynamic coefficient fitting parameters

$S(\cdot)$ = logistic function

$\phi$ = location of logistic curve halfway point

$m$ = logistic curve gradient parameter

$M$ = width of one-sided Gaussian

$w$ = height of one-sided Gaussian

$k$ = normalised parameter for elevator deflection

$\kappa$ = local reduced frequency

$r$ = local reduced pitch rate

$\Omega$ = local frequency of aerofoil motion, rad/s

$F_{\text{prop}}$ = thrust force, N

T/W = thrust-to-weight ratio of UAV

$\mathbf{z}$ = first-order system state

$\boldsymbol{\upsilon}$ = control input vector

$\mathbf{p}$ = vector of mixing parameters

$\boldsymbol{\alpha}$ = vector of angles of attack

$\mathrm{B}_i(\cdot)$, $\mathrm{T}_i(\cdot)$ = first-order system matrix functionals

$\mathbf{f}(\cdot)$ = first-order system vector function

$\boldsymbol{Q}_{\mathrm{A}}(\cdot)$ = aerodynamic model function

I = identity matrix

$[\cdot]_{\mathrm{IV}}$ = Iverson bracket

**Subscripts:**

$l$ = leading edge of aerofoil

$t$ = trailing edge of aerofoil

$L$ = lift

$D$ = drag

$M$ = pitching moment

$\alpha$ = coefficient with respect to angle of attack

sep = separated flow

att = attached flow

sym = symmetric component of model

ref = reference value

# 1. Introduction

Birds, bats and other flying animals show manoeuvrability far beyond the performance of current unmanned aerial vehicles (UAVs) of comparable scale. Manoeuvres such as stall turns [1–3], whiffling [4,5], zero-airspeed rolling [6] and ballistic braking [7–9] are enabled by complex wing morphing, and defy the performance limits of conventional UAVs. In addition, displays of aggression (agonistic behaviour) and predator-prey interaction by these animals involve close-quarters body reorientations [10], extreme evasive manoeuvres [11], and swarm coordination [12]. Many flying animals could indeed be characterized as *supermanoeuvrable*, as per by Herbst [13] and Gal-Or [14]: showing controlled post-stall manoeuvrability, and the ability to rapidly reorient themselves independent of their flight trajectory. However, existing supermanoeuvrable aircraft and UAVs have derived their capability not from bioinspiration, but from advances in thrust vectoring technology [14,15] and the study of unstable airframes [16,17]. Biomimetic supermanoeuvrability has remained an understudied topic until recently [18–20], but with increasing interest in autonomous dogfighting UAVs [21,22], enhancing UAV manoeuvrability is increasingly relevant.

Biomechanical studies have revealed several distinct mechanisms by which animals achieve extreme flight manoeuvrability. At smaller spatial scales, including in insects and hummingbirds, manoeuvrability is typically achieved by thrust vectoring: altering the kinematics of a high-frequency wingbeat which also provides the majority of the lift force required to maintain flight [23,24]. As such, this mechanism of manoeuvrability requires flapping-wing propulsion, and cannot be isolated from it. However, at larger spatial scales, it is possible to partially distinguish between flapping-wing propulsion and morphing-wing supermanoeuvrability, as separate but related phenomena. For instance, flying squirrels, which are without any form of propulsion, show capability for supermaneuvers such as stall turning and ballistic braking [9,25–27]. Several species of birds can carry out zero-airspeed rolling and perching manoeuvres without flapping motion [6,28], and perform other manoeuvres such as stall turning in timescales under a single wingbeat cycle [29]. Aldridge [30] concluded from a lift coefficient analysis that several species of bats turn without beating their wings, and evolutionary studies indicate that in the lineage of birds, flight manoeuvrability evolved before a strong power stroke [31]. Together, these results suggest that, even without flapping-wing propulsion, wing morphing can enable forms of extreme post-stall manoeuvrability.

A pertinent question is thus whether biomimetic and/or conventional forms of supermanoeuvrability could be achieved in a biomimetic morphing-wing UAV. Within existing literature on biomimetic morphing-wing control, there is an emphasis on small morphing motions over conventional or near-conventional flight envelopes (lower angles-of-attack), with the goal of providing improved alternatives to conventional control surfaces [32,33]—including in efficiency [34], noise [35], and control authority [32,36,37]. Other studies in configuration morphing have explored biomimetic planform changes to suit different flight regimes (mission morphing) [38,39]. In this work, we demonstrate in simulation that there is untapped potential for a third form of biomimetic morphing: larger morphing motions enabling strongly transient post-stall manoeuvring. We provide case study simulations of a biomimetic morphing-wing UAV, of scale *c.* 1 m, and equipped with both a conventional propulsion system, and six degree-of-freedom (6-DOF) wing rotation control (asymmetric sweep, dihedral and incidence). We develop and integrate state-of-the-art Goman-Khrabrov dynamic stall models into the multibody dynamics of this UAV, and design a novel open-look guidance strategy for generating post-stall pitching manoeuvres. Using this strategy, we study two key forms of pitch-axis supermanoeuvrability: a rapid nose-pointing-and-shooting (RaNPAS) manoeuvre, a classical supermaneuver available to many high-performance jet aircraft [14,17]; and a biomimetic ballistic transition manoeuvre, used by several flying and gliding animals for landing on vertical walls [7–9]. In both cases, wing morphing allows these manoeuvres to be performed in a UAV which is flight-dynamically stable at the level-flight position; and which is equipped with a thrust-to-weight ratio of only 0.5. This demonstrates the potential of biomimetic morphing to enable both classical supermanoeuvrability as well as extreme manoeuvres derived directly from biological behaviour. In addition, the open-loop guidance strategy we develop provides a basis for future exploration more complex forms of yaw-axis and multi-axis supermanoeuvrability in this class of UAVs.

## 2. UAV platform and modelling framework

### 2.1. Case study platform

As a case study, we consider the biomimetic morphing-wing UAV platform described in [18], of scale *c.* 1 m. The platform is equipped with 6-DOF wing morphing: asymmetric sweep, dihedral and incidence. This is a maximally-actuated configuration to allow significant DOFs to be identified over the course of this study. It is also equipped with a generic axial thruster for propulsion. Airframe parameters are presented in Table 1, and a scale rendering in Fig. 1A, indicating the active morphing DOFs. The platform is approximately equivalent in scale to several larger birds (*e.g.*, the Greylag goose, *Anser anser*) as well as several existing morphing-wing uncrewed aerial vehicles (*e.g.*, the NextGen MFX-1). It also is deliberately designed to be conservative: wing masses are comparatively large, to account for wing strengthening to allow high angle-of-attack manoeuvring, and lifting surface chords are comparatively small. Figure 1B illustrates a hypothetical RaNPAS manoeuvre in this platform, with applications in dogfighting UAVs, *cf.* [14].

### 2.2. Flight dynamic modelling framework

To model the flight dynamics of this case study UAV under strongly transient post-stall manoeuvres, we extend and enhance a previous flight dynamic modelling framework for this UAV [18,40]. This previous framework modelled the UAV via a 12-DOF nonlinear state-space representation in the form:

$$\mathrm{B}_1(\mathbf{z},\boldsymbol{\upsilon})\dot{\mathbf{z}} = \mathbf{f}(\mathbf{z},\boldsymbol{\upsilon}) - \mathrm{B}_0(\mathbf{z},\boldsymbol{\upsilon})\mathbf{z}, \quad (1)$$

with a set of morphing and control parameters $\boldsymbol{\upsilon}$; a state variable $\mathbf{z} \in \mathbb{R}^{12}$ which encodes the reference rotational and translational positions and velocities of the system; and functionals

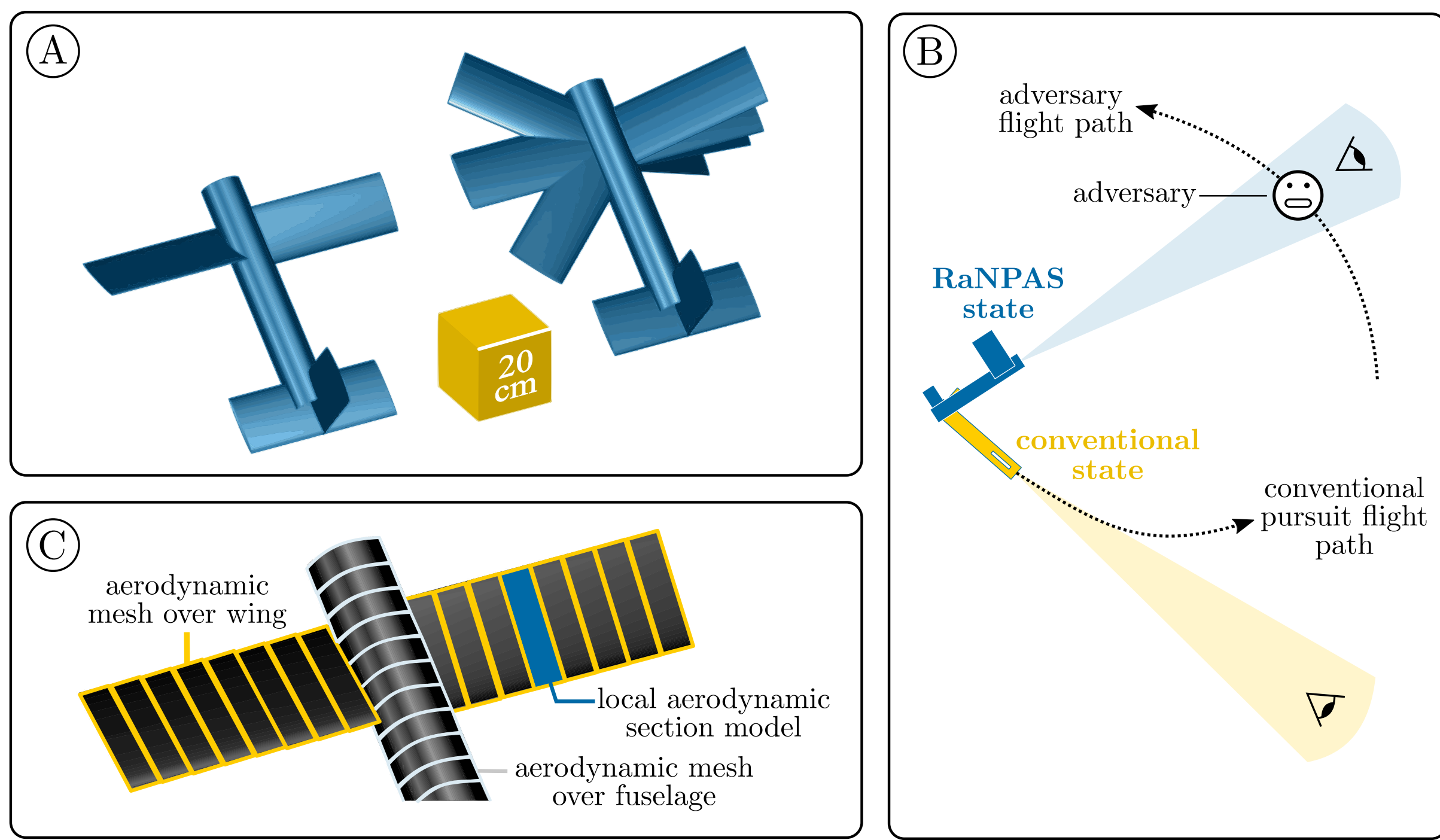

**Figure 1:** Illustration of the case study biomimetic morphing-wing UAV. (**A**) Morphing degrees of freedom of the case study system: wing incidence, sweep, and dihedral, all independently controllable on both wings. (**B**) Dogfighting context of a RaNPAS manoeuvre: the ability to significantly alter the UAV field of view, independent of the flight path. (**C**) An illustrative mesh of aerodynamic section models for the UAV lifting surface and fuselage.

**Table 1:** Hybrid system properties with comparisons: n/a, and n/spec denote data not available and not relevant to be specified, respectively.

| | This study | NextGen MFX-1 [41] | ShowTime 50 [42] | Greylag Goose (*A. Anser*) [43–45] |
|---|---|---|---|---|
| Properties: | Values: | | | |
| Length – fuselage | 1.20 m | 2.1 m | 1.51 m | *c.* 0.82 m |
| Length – wing to tail | 0.80 m | *c.* 1.17 m | *c.* 0.94 m | n/a |
| Length – body radius | 0.10 m | *c.* 0.15 m | *c.* 0.088 m | n/a |
| Span – wing | 1.60 m | 2.8 m | 1.46 m | *c.* 1.62 m |
| Span – horz. stabilizer | 0.80 m | n/a | *c.* 0.62 m | n/a |
| Span – vert. stabilizer | 0.40 m | n/a | *c.* 0.17 m | n/a |
| Chord – wing | 0.15 m | *c.* 0.23 | 0.32 m | *c.* 0.26 m |
| Chord – horz. stabilizer | 0.15 m | n/a | *c.* 0.22 m | n/a |
| Chord – vert. stabilizer | 0.15 m | n/a | *c.* 0.09 m | n/a |
| Aerofoils | ST50 W/H | n/a | ST50 W/H/V | Complex |
| Mass – total | 8 kg | 45 kg | 2.9 kg | *c.* 3.3 kg |
| Mass – single wing | 1 kg | n/a | n/a | n/a |
| Propulsion – max. thrust | n/spec | *c.* 200 N | *c.* 60 N | n/a |
| Propulsion – mechanism | n/spec | jet engine | propeller | flapping-wing |

$\mathrm{B}_1$, $\mathrm{B}_0$, and $\mathbf{f}$ describing the UAV flight dynamics. These functionals account for the multibody dynamics arising from wing morphing, under the assumption of ideal actuation—detailed definitions are given in the Supplementary material. Within $\mathbf{z}$, the UAV's orientation is encoded in Euler angles: the gimbal lock problem at the Euler angle pole is bypassed via a pole-switching routine, in which two alternate Euler angle parameterizations maintain non-singularity over the complete orientation space [46]. The aerodynamics of the UAV are modelled via a mesh of local aerodynamic section models over each of its lifting surfaces, and the fuselage (Fig. 1C). The dependency of the UAV multibody dynamic model (Eq. 12) on this local aerodynamic model, $\boldsymbol{Q}_\mathrm{A}(\mathbf{z},\boldsymbol{\upsilon})$, may be represented as:

$$\mathrm{B}_1\big(\mathbf{z},\boldsymbol{\upsilon},\boldsymbol{Q}_\mathbf{A}(\mathbf{z},\boldsymbol{\upsilon})\big)\dot{\mathbf{z}} = \mathbf{f}\big(\mathbf{z},\boldsymbol{\upsilon},\boldsymbol{Q}_\mathbf{A}(\mathbf{z},\boldsymbol{\upsilon})\big) - \mathrm{B}_0\big(\mathbf{z},\boldsymbol{\upsilon},\boldsymbol{Q}_\mathbf{A}(\mathbf{z},\boldsymbol{\upsilon})\big)\mathbf{z}. \quad (2)$$

Previous work [18] developed a quasisteady aerodynamic model for $\boldsymbol{Q}_\mathrm{A}(\mathbf{z},\boldsymbol{\upsilon})$, and validated this model against empirical data for UAV flight dynamics at low aerodynamic unsteadiness. However, to model strongly unsteady RaNPAS behaviour with sufficient accuracy, a quasisteady aerodynamic model is not sufficient [47,48]. For UAV flight simulation under strongly unsteady flow (local reduced frequency $0.01 < \kappa < 0.5$), several approaches are established: (**i**) fully three-dimensional computational fluid dynamics (CFD) simulations [49,50]; (**ii**) phenomenological dynamic stall and lift hysteresis models, such as the Goman-Khrabrov (GK) [51] model, integrated into the section model framework [47,48,52–54]; and (**iii**) model-reduction and machine-learning (ML) techniques [55–57]—applied to higher-fidelity data to generate an accurate surrogate aerodynamic model. Here, we utilize a GK dynamic stall model, accounting for strongly transient effects arising from aerofoil pitching motion during the pitching manoeuvres we will study. GK models have been previously utilized in the study of other agile and morphing-wing UAVs [48,53]. In §3, we detail the construction of a modified GK model for this UAV.

## 3. Goman-Khrabrov (GK) aerodynamic modelling

### 3.1. GK model formulation

To account for the transient effects of UAV pitching motion, we implement a novel Goman-Khrabrov (GK) model into the flight dynamic framework of §2.2, extending both state-of-the-art GK models [47,53], and the quasisteady flight dynamic model of this UAV [18]. Following state-of-the-art GK models, for each lifting surface station model within the aerodynamic mesh, the aerodynamic coefficients for force or moment $i$ ($C_i$) as a function of effective angle-of-attack ($\alpha$) are defined by the mixing function:

$$C_i(\alpha) = pC_{i,\text{att}}(\alpha) + (1-p)C_{i,\text{sep}}(\alpha), \tag{3}$$

where $C_{i,\text{att}}(\alpha)$ and $C_{i,\text{sep}}(\alpha)$ are the aerodynamic coefficient functions for hypothetical cases of local attached and separated flow respectively. Note that Eq. 3 applies independently for each lifting surface station model; but for readability we will omit indexing across each station. The three forces ($i$) of relevance are lift ($L$), drag ($D$), and pitching moment ($M$). Parameters $p_i$ are local dynamic mixing parameters, loosely connected to the location of the separation point along the aerofoil chord [58], and governed by the first-order differential equation:

$$\tau_1\dot{p}(\alpha) = p_0(\alpha - \tau_2\dot{\alpha}) - p\alpha, \tag{4}$$

where $\alpha$ and $\dot{\alpha}$ are the local angle of attack and corresponding rate; $\tau_1$ and $\tau_2$ are delay parameters; and $p_0(\alpha)$ are mixing functions representing the transition between attached and separated flow. To implement this model, we must identify the quasistatic functions $C_{i,\text{att}}(\alpha)$, $C_{i,\text{sep}}(\alpha)$ and $p_0(\alpha)$ and the transient delay parameters $\tau_1$ and $\tau_2$ for the cast study UAV aerofoils (Table 1). The three quasistatic functions are identifiable based only on quasisteady aerodynamic coefficient data (when $p = p_0$), whereas the transient delay parameters are only identifiable with transient aerodynamic data.

### 3.2. Wing parameter identification

Quasisteady aerodynamic data for the two aerofoils (ST50W, ST50H) in the case study UAV have been previously obtained by Selig [59] (Fig. 2), and these data permit the identification of quasistatic GK model functions. Beginning with the wing aerofoil: nonlinear least-squares curve fitting indicates that it is poorly approximated by the flat-plate aerofoil models that are traditionally used in the GK modelling [48,54], but well-approximated by extended models [58,60,61]. For separated flow, we use:

$$\begin{aligned}
C_{L,\mathrm{sep}}(\alpha) &= a_L \operatorname{sgn}\alpha \sin(b_L|\alpha + c_L| + d_L) + e_L, \\
C_{D,\mathrm{sep}}(\alpha) &= a_D \sin(b_D|\alpha| + c_D) + d_D, \\
C_{M,\mathrm{sep}}(\alpha) &= a_M \operatorname{sgn}\alpha \sin(b_M|\alpha + c_M| + d_M) + e_M,
\end{aligned} \tag{5}$$

for all $\alpha$, with model parameters $a_i$- $e_i$, and where sgn· is the signum function. For attached flow, the difference in geometry between the aerofoil leading and trailing edge necessitates that we treat these edges separately:

$$\begin{aligned}
C_{L,\mathrm{att},l}(\alpha_l) &= C_{L\alpha,l}\alpha_l, & C_{L,\mathrm{att},t}(\alpha_t) &= C_{L\alpha,t}\alpha_t, \\
C_{M,\mathrm{att},l}(\alpha_l) &= 0, & C_{M,\mathrm{att},t}(\alpha_t) &= C_{M\alpha,t}\alpha_t, \\
C_{D,\mathrm{att},l}(\alpha_l) &= 0, & C_{D,\mathrm{att},t}(\alpha_t) &= 0,
\end{aligned} \tag{6}$$

where $\alpha_l$ and $\alpha_t$ are the leading and trailing edge angles of attack, representing a partition of the full domain, $|\alpha| < 180°$, into $|\alpha| \leq 90°$ and $\big||\alpha| - 180°\big| \leq 90°$, the latter of which is mapped back to $|\alpha| \leq 90°$ again. Three of the attached flow models are observably zero (Eq. 6, Fig. 2). The effect of aileron deflection is not considered, as this control function will be achieved by incidence morphing.

Model parameters for Eq. 5-6 are identified via a nonlinear least-squares approach applied to selections of clearly attached and separated flow. Data-driven estimates of $p_0$ can then be obtained by solving $p = p_0$ in Eq. 3 using aerodynamic source data. Figure 3 shows the results of this process: for the leading edge ($\alpha_l$), compared to the compared to a traditional arctangent expression for $p_0(\alpha)$, as per Wickenheiser and Garcia [54] and Reich et al. [48]:

$$p_{0,l}(\alpha_l) = \begin{cases} 1 & |\alpha_l| < 7° \\ -0.3326\tan^{-1}(|\alpha_l| + 16) + 0.5 & 7° \leq |\alpha_l| \leq 37° \\ 0 & |\alpha_l| > 37°. \end{cases} \tag{7}$$

Note that $\tan^{-1}(\cdot)$ is here taken to output a value in radians. For the trailing edge, we modify

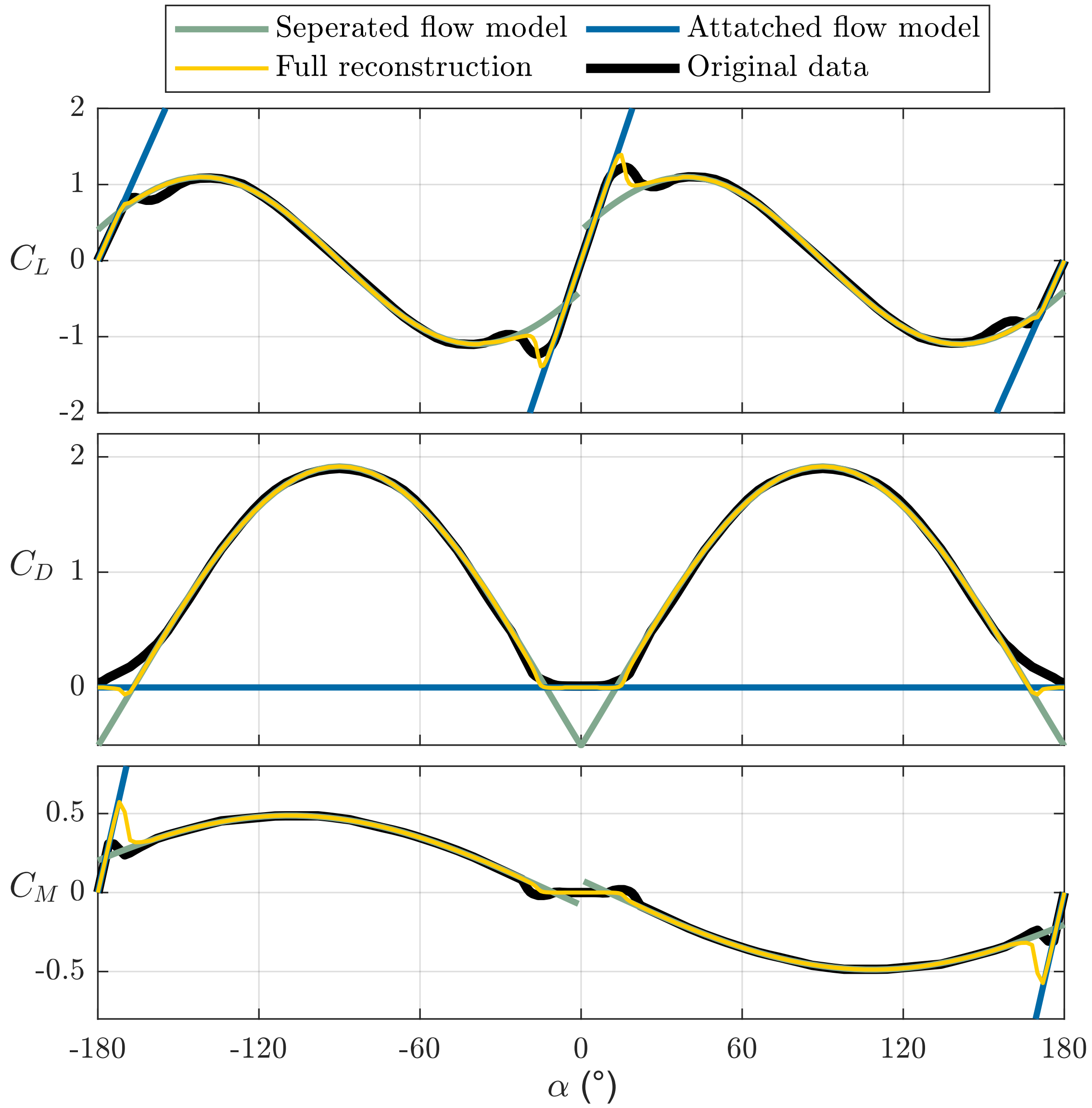

**Figure 2:** Quasisteady aerodynamic coefficient data for the wing aerofoil (ST50W), as a function of angle-of-attack ($\alpha$), reconstructed from the quasistatic GK attached and separated flow models, compared to the original data.

this expression to account for earlier and faster separation:

$$p_{0,t}(\alpha_t) = \begin{cases} 1 & |\alpha_t| < 4^\circ \\ -0.3326\tan^{-1}(1.6|\alpha_t| + 16) + 0.5 & 4^\circ \leq |\alpha_t| \leq 21^\circ \\ 0 & |\alpha_t| > 21^\circ. \end{cases} \tag{8}$$

Quasisteady aerodynamic coefficient data for the ST50W wing may then be reconstructed for comparison. Figure 2 shows this data alongside the GK reconstruction using the arctangent $p_0$ (Eq. 7-8). The result is overall very good: the separated and attached flow regimes are modelled well. Discrepancies are observed in trailing edge transition in drag and moment coefficients: as can be seen in Fig. 3, trailing edge drag and moment appear to behave according to different $p_0(\alpha)$ functions. However, defining three independent mixing functions would be overfitting these data, and would break the physical interpretability of $p$ as a mixing parameter.

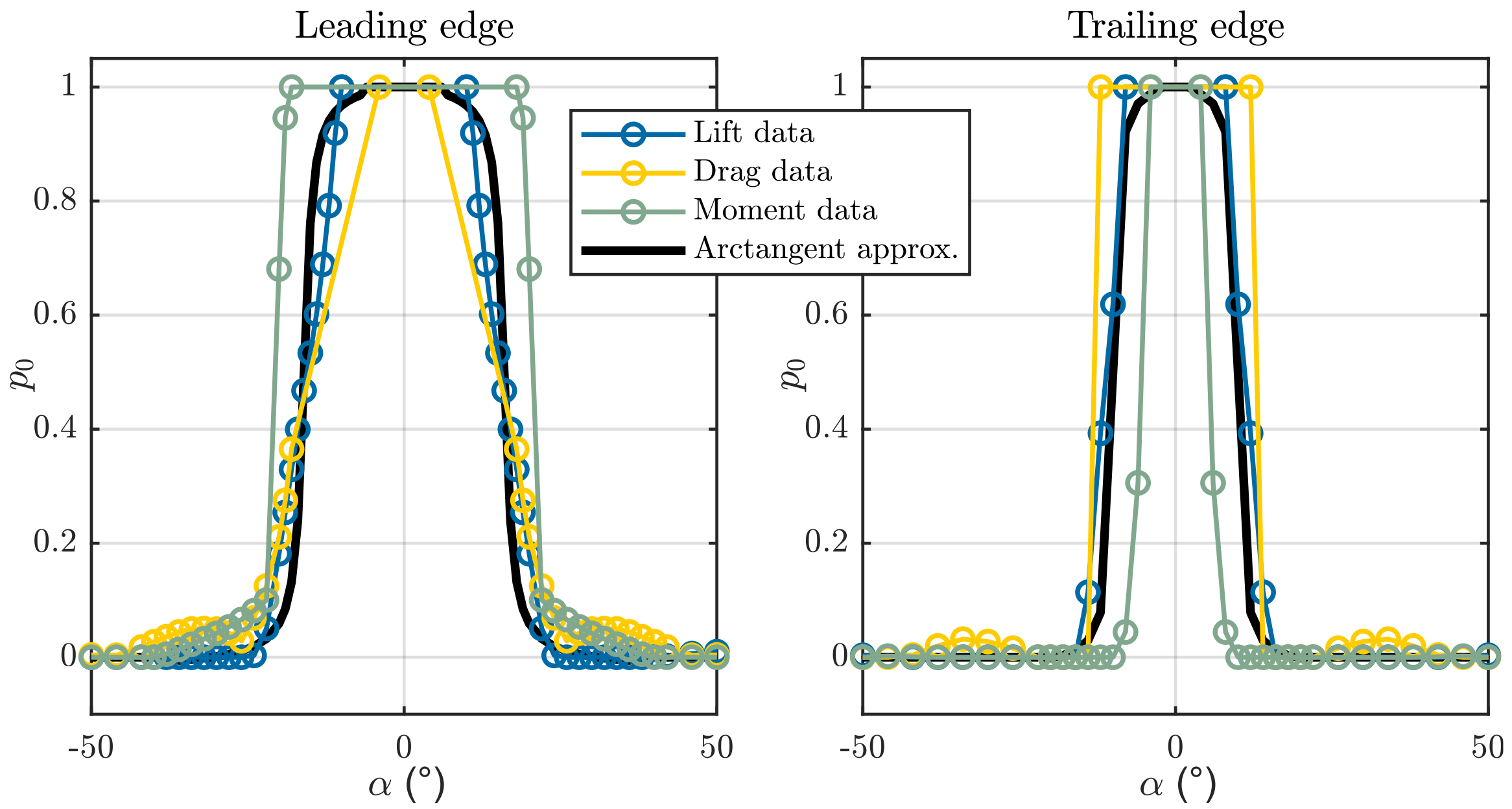

**Figure 3:** Data-driven estimates of $p_0(\alpha)$ derived from wing aerofoil (ST50W) leading and trailing edge aerodynamic data, compared to arctangent approximations (Eq. 7-8).

### 3.3. Stabiliser parameter identification

The aerodynamic data for the stabilizer aerofoil (ST50H) [59], is dependent on the stabiliser control surface (elevator/rudder) deflection. To begin, we assume that control-surface motion can be modelled quasistatically (*i.e.*, that this motion induces no flow). The dataset from Selig [59] contains aerodynamic coefficient data at seven different elevator deflections ($-50°$, $-30°$, $-15°$, $0°$, $15°$, $30°$, $50°$); but only four are unique (*e.g.*, $\beta_e \in [-50, 0]°$) due to the symmetric aerofoil profile: downwards aerofoil motion at downwards control surface deflection is equivalent to upwards motion at upwards deflection. For each element of the unique set $\beta_e \in [-50, 0]°$, we identify separated flow models of the form:

$$
\begin{aligned}
C_{L,\text{sep}}(\alpha) &= a_L \operatorname{sgn}(\alpha + c_L) \sin(b_L|\alpha + c_L| + d_L) + e_L, \\
C_{D,\text{sep}}(\alpha) &= \begin{cases} a_D \cos(b_D|\alpha + c_D| + d_D) + e_D, & \beta_e = 0 \\ a_D \sin(b_D \alpha + c_D) + d_D, & \text{o.w.}, \end{cases} \\
C_{M,\text{sep}}(\alpha) &= a_M \operatorname{sgn}(\alpha + c_M) \sin(b_M|\alpha + c_M| + d_M) + e_M,
\end{aligned} \tag{9}
$$

and attached-flow models as per Eq. 6. The key difference between stabilizer (Eq. 9) and wing (Eq. 5) models is the simpler sinusoid drag model for the stabilizer at nonzero $\beta_e$: the complexity of the coefficient data does not permit identification of more complex models. We use a nonlinear least-squares approach to identify model parameters for Eq. 6 across the unique $\beta_e$. This identification is fully automated except for a manual indication of the location of areas of attached and separated flow for identification. The Supplementary Material presents the four unique identified models in each aerodynamic coefficient.

We then identify the mixing parameter functions, $p_0(\alpha)$. Figure 4 shows data-driven estimates of $p_0(\alpha)$ obtained by solving Eq. 3 for $p = p_0$ in the vicinity of transition. Estimates are available for $C_L$ at the leading and trailing edge, and $C_M$ at the trailing edge—areas where the attached flow model is nonzero. In Fig. 4, these estimates are presented with respect to the reference angles-of-attack, $\alpha_{l,\mathrm{ref}}$ and $\alpha_{t,\mathrm{ref}}$: these values are the centre-points of the attached flow regions, specified manually, and nonzero for nonzero $\beta_e$. A notable feature of these results is their asymmetry, with long tails at negative $\alpha$ (for $\beta_e < 0$). This is likely a physical effect. At positive $\alpha$ values (for $\beta_e < 0$), large stall peaks are observed, whereas at negative $\alpha$ there is a flat plateau: physically, this could arise from flow reattachment effects when both the control surface and the aerofoil are inclined upwards ($\beta_e < 0$, $\alpha > 0$), leading to a state in which the control surface is itself at lower angle-of-attack. The arctangent sigmoid used in previous GK models [48] cannot capture this asymmetry. In its place we propose a new GK sigmoid function, based on the logistic function. Its symmetric form, for the leading edge ($\alpha_l$), is:

$$p_{0,l,\mathrm{sym}}(\alpha_l) = S\left(\frac{1}{m_l}\left(\left|\alpha_l - \alpha_{l,\mathrm{ref}}\right| - \phi_l\right)\right), \qquad S(x) = \frac{1}{1+\exp(x)}, \tag{10}$$

where $S(x)$ is the logistic function and $\alpha_{l,\mathrm{ref}}$ is the centre point of the attached flow region (specified manually). The shift parameter $\phi_l$ is the location of the halfway point, *i.e.*, $p_{0,l,\mathrm{sym}}(\phi_l + \alpha_{l,\mathrm{ref}}) = 0.5$. The width parameter $m_l$ governs the gradient at this point. The interpretable nature of these parameters is an aid to identification. To account for asymmetry in angle-of-attack, we add a one-sided Gaussian term to Eq. 10, yielding the completed $p_{0,l}$:

$$\begin{aligned} p_{0,l}(\alpha_l) &= \left(1 - p_{0,\mathrm{sym}}(\alpha_l)\right) G(\alpha_l) + p_{0,i,\mathrm{sym}}(\alpha_l), \\ G(\alpha_l) &= M_l \exp\left(-\left(\frac{\alpha_l - \alpha_{l,\mathrm{ref}} + \phi_l}{w_l}\right)^2\right)\left[\alpha_l - \alpha_{l,\mathrm{ref}} < 0\right]_{\mathrm{IV}}, \end{aligned} \tag{11}$$

where $w_l$ governs the width of the one-sided Gaussian function, and the parameter $M_l$ its height. $\phi_l$ is the parameter identified in Eq. 10, and $[\cdot]_{\mathrm{IV}}$ is the Iverson bracket [62], such that $[s]_{\mathrm{IV}} = 1$ if $s$ is true, and $[s]_{\mathrm{IV}} = 0$ if $s$ is false. This addition maintains smoothness ($C^\infty$) over the halfspaces $\alpha_l > \alpha_{l,\mathrm{ref}}$ and $\alpha_l < \alpha_{l,\mathrm{ref}}$.

In the case of the trailing edge ($\alpha_t$), discrepancy between the empirical $p_0$ estimates computed from $C_L$ and $C_M$ precludes identification of an asymmetric $p_{0,t}$. We instead use the same symmetric form as in Eq. 10, with parameters $m_t$ and $\phi_t$:

$$p_{0,t,\mathrm{sym}}(\alpha_t) = S\left(\frac{1}{m_t}\left(\left|\alpha_t - \alpha_{t,\mathrm{ref}}\right| - \phi_t\right)\right), \tag{12}$$

For identification, all $p_0$ parameters (for both leading and trailing edge) are manually estimated for $\beta_e = -50°$ and $\beta_e = 0°$; and models at the internal surface-deflection points are generated by linear interpolation. Table 2 shows the identified parameters, including the interpolation index ($k \in [0, 1]$ for $\beta_e \in [-50, 0]°$), and Fig. 4 the identified $p_0(\alpha)$ functions. The parameter interpolation is linear and two-point ($k \in \{0, 1\}$), with the exception of $M_l$, which shows a rising trend with $k$ but must be zero at $k = 1$ to preserve symmetry. To account for this effect, we use a non-monotonic piecewise-linear profile (Table 2). For both edges, the complete set of identified models can be extended to $\beta_e > 0$ by symmetry, and estimated quasisteady coefficient profiles can be reconstructed using the relevant sigmoid $p_0$ expressions and the separated- and attached-flow models. Figure 5 shows the GK reconstruction of the ST50H quasisteady aerodynamic coefficients as a function of elevator deflection and $\alpha$, compared with the original results of Selig [59]. As can be seen, a good agreement is observed, despite some variation in the laminar-turbulent transition zones. The primary limitations of the model remain the discrepancy in identified separation point between the lift and moment coefficient data.

**Table 2:** Fitted model parameters for the logistic $p_0$ functions

| Parameter | $\beta_e = -50°$ | $\beta_e = -30°$ | $\beta_e = -15°$ | $\beta_e = 0°$ |
|---|---|---|---|---|
| $k$ | 0 | 0.4 | 0.7 | 1 |
| $\alpha_{l,\mathrm{ref}}$ | 30.5° | 19° | 8° | 0° |
| $\alpha_{t,\mathrm{ref}}$ | −148° | −162° | −172° | −180° |
| $m_l$ | 0.8 | lin. interp. | | 3 |
| $\phi_l$ | 4.4 | lin. interp. | | 20 |
| $M_l$ | 0.4 | lin. interp. via 0.6 at 0.75$k$ | | 0 |
| $w_l$ | 18 | lin. interp. | | 14 |
| $m_t$ | 1.5 | constant | | 1.5 |
| $\phi_t$ | 11 | lin. interp. | | 9 |

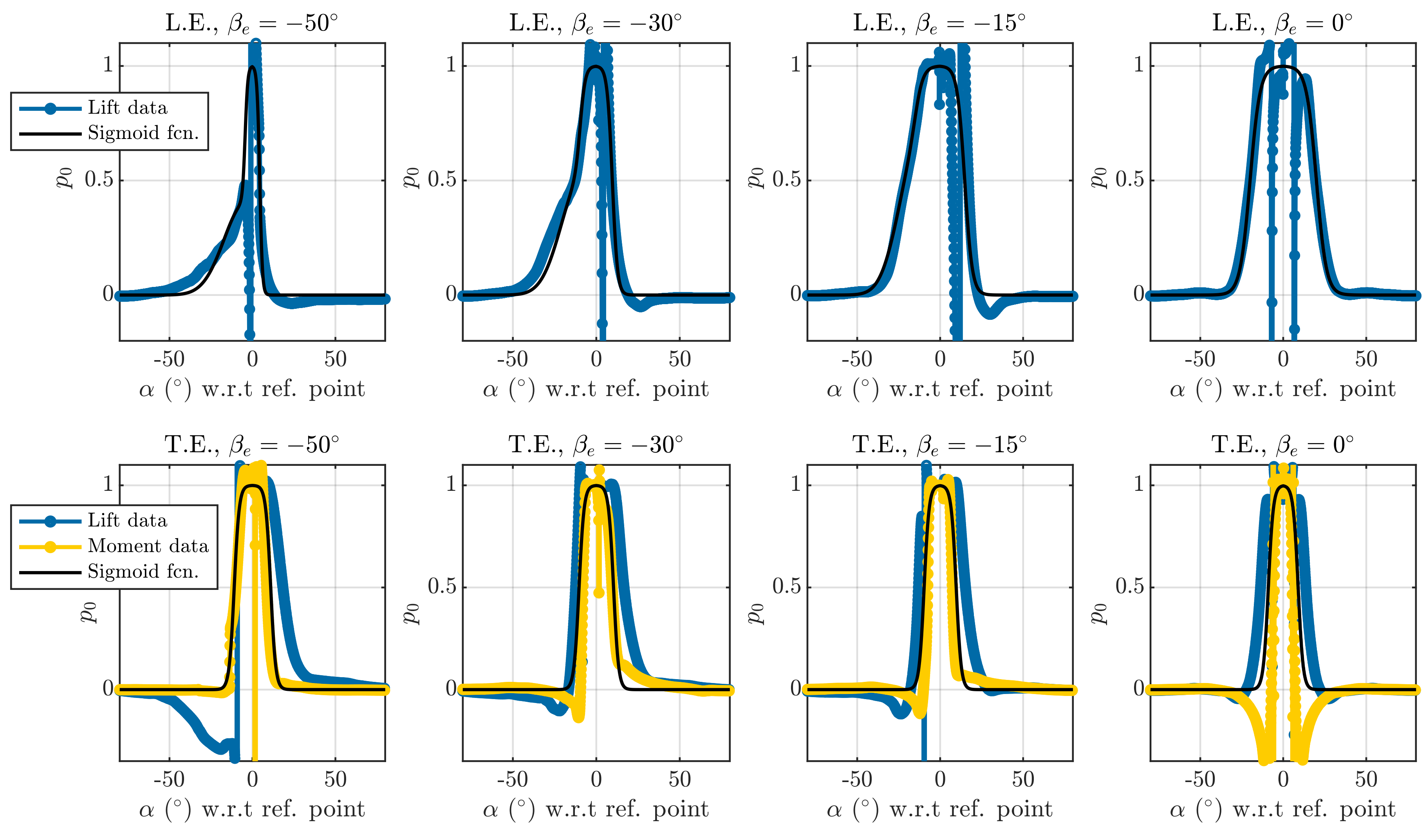

**Figure 4:** Unfiltered approximations to $p_0(\alpha)$ derived from stabiliser aerofoil (ST50H) leading edge (L.E.) and trailing edge (T.E.) aerodynamic data, against the associated logistic sigmoid fit.

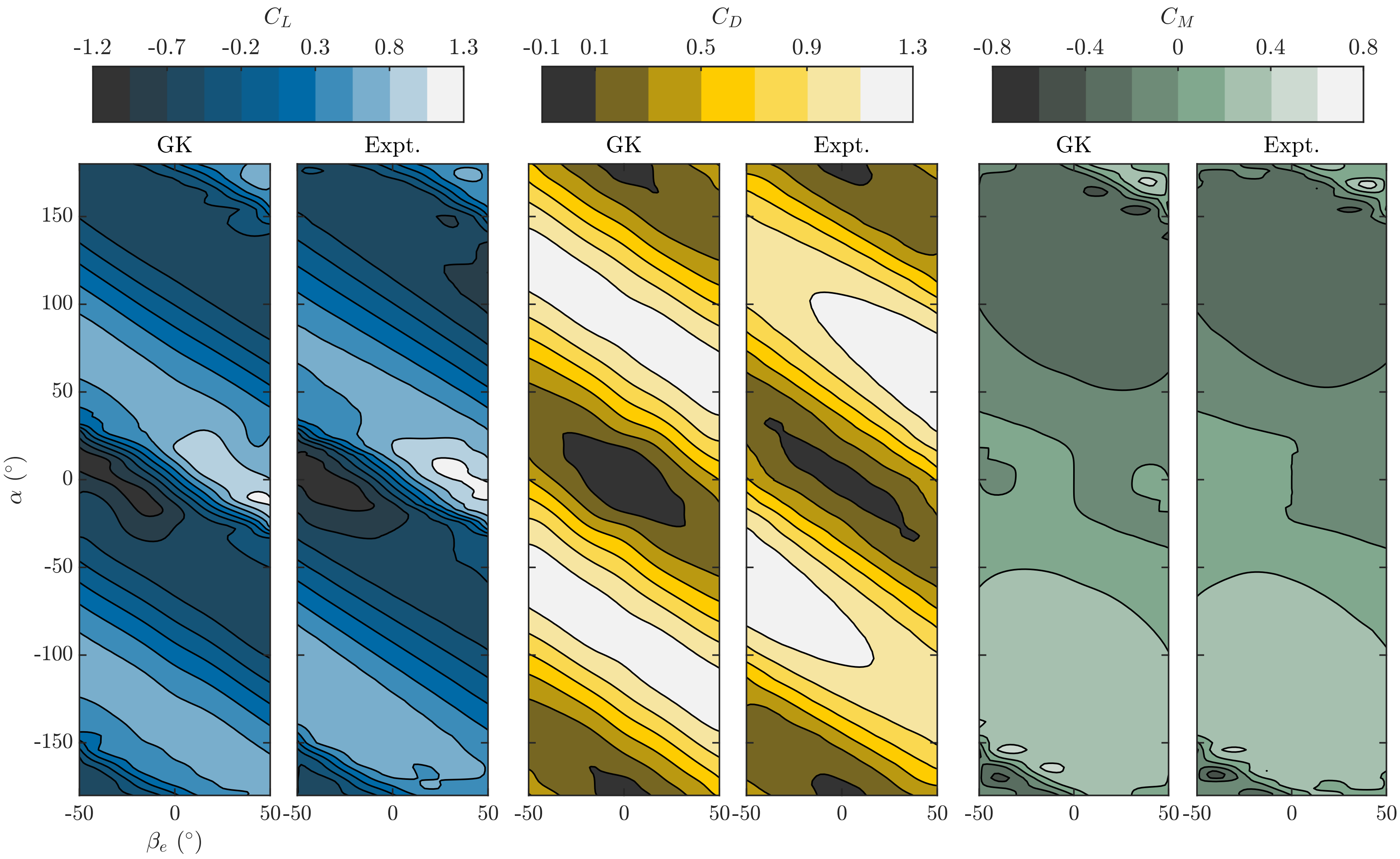

**Figure 5:** Quasisteady aerodynamic coefficient data for the stabiliser aerofoil (ST50H), as a function of both angle-of-attack ($\alpha$) and control surface deflection ($\beta_e$), reconstructed from the quasistatic GK attached and separated flow models, and compared to the original data.

### 3.4. Transient delay parameter identification

With the quasistatic components of the GK models for both UAV aerofoils fully identified, the remaining task is to identify the transient delay parameters $\tau_1$ and $\tau_2$. For comparison, Table 3 presents a range of delay parameters identified in the literature for three different aerofoils. The variation across reported values is large: for the NACA0009 factor of 2 variation (in $\tau_2$) is observed across the reported values; for the NACA0012 a factor of 4 (in $\tau_1$); and for the NACA0018 a factor of 7 (in $\tau_1$). These results indicate that a precise identification of the delay parameters is sensitive to the dataset—contributing factors could include wind-tunnel/wall effects, surface roughness, and CFD modelling inaccuracies. Additionally, this is consistent with the observation that these delay parameters determine the aerofoil behaviour in the laminar-turbulent transition, including the case of attached flow at angles-of-attack below quasisteady stall: both factors are strongly dependent on modelling and dataset specifics.

**Table 3:** GK delay parameters reported in the literature

| Aerofoil | $\tau_1$ $(c/U)$ | $\tau_2$ $(c/U)$ | Source | Reynolds No. |
|---|---|---|---|---|
| NACA0009 | 2.5 | 2.0 | An et al. [63] | $4.9 \times 10^4$ |
| | 2.28 | 3.41 | Reich et al. [48] | not stated |
| | 2.6 | 2.8 | Williams et al. [58] | $5 \times 10^4$ |
| | 3.6 | 4.3 | Williams et al. [60] | $5.7 \times 10^4$ |
| NACA0015 | 0.52 | 4.5 | Goman [64] | c. $2 \times 10^5$ |
| | 2.14 | 13.56 | Fan [65] | not stated |
| NACA0018 | 0.25 | 3.6 | Williams et al. [58] | $2.5 \times 10^5$ |
| | 1.57 | 1.52 | Greenblatt et al. [66] | $3 \times 10^5$ |
| | 1.73 | 4.83 | Niel [67] | $1.9 \times 10^5$ |

A CFD process for modelling the ST50W aerofoil is described in [46]. From it we estimate $\tau_1 = \tau_2 = 2.3c/U$. This is broadly consistent with the NACA0009 estimates given in Table 3. Two upper bounds on this estimate, in terms of motion transience, are defined. The maximum permissible reduced frequency is $\kappa = b\Omega/U = 0.5$, where $b$ is the local section semichord, and $\Omega$ is the frequency (in rad/s) of aerofoil motion. The maximum permissible reduced pitch rate is $r = b\dot{\alpha}/U = 0.13$, where $\dot{\alpha}$ is the local section pitch rate. We note that there are several open questions with regard to how to integrate GK dynamic stall models into a flight simulation context [46]: most significantly, exactly how the delay parameters should be taken to scale with local station airspeed ($U$), given that this airspeed will vary. We take these delay parameters to scale with $U$, according to the dimensional relation $\tau_1 = \tau_2 = 2.3c/U$, but further research is required to establish this relationship with confidence.

### 3.5. Combined multibody-GK model framework

The completed GK model defines the aerodynamic force function $Q_{\mathrm{A}}(\mathrm{z},\upsilon)$ in the system dynamics (Eq. 2). The combination of GK and multibody-dynamic models for the case-study UAV lead to the nonlinear state-space model:

$$\begin{bmatrix} \mathbf{B_1}(\mathbf{z},\boldsymbol{\upsilon}) & \\ & \mathrm{T}_1(\mathbf{z},\boldsymbol{\upsilon}) \end{bmatrix} \begin{bmatrix} \dot{\mathbf{z}} \\ \dot{\mathbf{p}} \end{bmatrix} = \begin{bmatrix} \mathbf{f}(\mathbf{z},\mathbf{p},\boldsymbol{\upsilon}) \\ \boldsymbol{p_0}\big(\boldsymbol{\alpha}(\mathbf{z},\boldsymbol{\upsilon}) - \mathrm{T}_2(\mathbf{z},\boldsymbol{\upsilon})\dot{\boldsymbol{\alpha}}(\mathbf{z},\boldsymbol{\upsilon})\big) \end{bmatrix} + \begin{bmatrix} -\mathrm{B}_0(\mathbf{z},\boldsymbol{\upsilon}) & \\ & \mathrm{I} \end{bmatrix} \begin{bmatrix} \mathbf{z} \\ \mathbf{p} \end{bmatrix}, \tag{13}$$

where the terms in $\mathbf{p}$ and $\dot{\mathbf{p}}$ represent the flow attachment dynamics (Eq. 4) over all lifting surfaces ($p$ now becoming $p_j$ for mesh station $j$), and the terms in $\mathbf{z}$ and $\dot{\mathbf{z}}$ represent the UAV multibody dynamics (Eq. 2). The addition of the flow attachment dynamics significantly increases the size of the state space. We perform a mesh independence study which indicates that five aerodynamic stations along each lifting surface ensure convergence in overall lifting surface lift, drag and moment to below 1% for the manoeuvres we will study (§4-6). This leads to an aerodynamic system ($\mathbf{p}$, $\dot{\mathbf{p}}$) with 25 degrees of freedom, in addition to the 12 degrees of freedom of the first-order multibody dynamics model ($\mathbf{z}$, $\dot{\mathbf{z}}$). We will now apply this flight dynamic model to understand the capability of the case study UAV for supermanoeuvrability.

## 4. Classical RaNPAS capabilities

### 4.1. Manoeuvre design

Gal-Or's [14] classification of rapid nose-pointing-and-shooting (RaNPAS) capability includes the supermanoeuvre commonly referred to as the *cobra*. This is a pitch-axis supermanoeuvre which involves tilting the UAV backwards from level flight to beyond 90° pitch angle, and then forwards to level flight again, while maintaining approximately constant altitude [17]. A range of supermanoeuvrable jet aircraft are capable of this manoeuvre [14,15], but it is not observed in animals—as such, it represents an initial test case of whether biomimetic mechanisms enable classical supermanoeuvrability. At the simplest level, recreating a cobra manoeuvre in the case study UAV requires three control configurations: (**1**) an initial trim configuration, (**2**) a configuration to generate the moment required to pitch the UAV up to the partially inverted position; and (**3**) a configuration to pitch the UAV down from the partially inverted position, and back to trim. The initial trim configuration (1) is computable via existing trimming methods for this UAV [18], and additionally may be a candidate for configuration (3). However, to identify configuration (2) as well as additional candidates for configuration (3), we develop a novel approach, as follows.

Firstly, as the cobra manoeuvre is constrained to the $x$-$z$ plane, the control space for manoeuvre design is constrained by symmetry about this plane. Available morphing degrees of freedom are the symmetric dihedral $\Gamma$, the symmetric sweep $\Lambda$, and the symmetric incidence $\Phi$. Other available control degrees of freedom are the elevator deflection $\beta_e$, and propulsive force $F_{\text{prop}}$. For physical feasibility, some of these degrees of freedom should be constrained: we enforce control limits on the elevator deflection ($|\beta_e| < 0.87$ rad, 50°) and wing sweep ($|\Lambda| < 1.171$ rad, 67°). Secondly, for the initial manoeuvre design phase we utilise the quasisteady aerodynamic model for this UAV [59]. The purpose of this initial model simplification is to permit a characterisation of the UAV's nonlinear longitudinal static stability characteristics, which we will optimise to generate candidate control configurations. Thirdly, in order to automatically identify candidate morphing configurations for the RaNPAS manoeuvre, we define objective functions related to the intended behaviour of the configuration. Multiple objective functions are available. For the pitch up configuration (2), one option is the UAV point pitch acceleration at a given angle of attack, $\ddot{\theta}(\theta)$. Others include the pitch acceleration integral ($\int \ddot{\theta}(\theta)\, d\theta$), and the location of the roots of the UAV's nonlinear longitudinal static stability profile: $\theta : \ddot{\theta}(\theta) = 0$. We refer to this root as a *quasi-trim* state: this state is momentarily at pitch equilibrium ($\ddot{\theta} = 0$). However, it is not at equilibrium in translational degrees of freedom (airspeed, or altitude), and so is eventually likely to deviate from an orientation equilibrium, as changes in airspeed and altitude / altitude rate propagate to changes in pitch dynamics. The process may be analogised with fast-slow behaviour in dynamical systems [68]. Quasi-trim states will be of significant relevant to our characterisation of pitch-axis supermanoeuvrability.

As an initial objective function for the pitch-up state (2), we use the point pitch acceleration at a pitch value of 0.8 rad (46°). Figure 6 shows several morphing configurations generated by nonlinear least-squares optimization. Configurations A-C indicate pitch stability plots for optimal configurations with (A) all degrees of freedom active, (B) sweep and incidence active and (C) only incidence active. The associated wing configuration is rendered alongside Note that the UAV airspeed is $U = 30$ m/s, and thrust is $F_{\text{prop}} = 10$ N, following [18,46] and *cf.* Table 1. In the case of (A), however, note that significant additional pitch-up moment can be generated by the offset between the propulsive force axis and the centre of mass due to the upwards wing dihedral. Parameter values for these configurations are given in Table 4; values in bold type are located on their respective constraint limits—indicating the effect of these constraints on the configuration performance. For example, in all states the elevator is at its control limit, and it is self-evident that increased elevator control effectiveness will result in greater pitch control effectiveness at low angle-of-attack. However, at very high angles-of-attack (> 1 rad) the elevator ceases to have a significant effect on the system pitch dynamics, and morphing controls must take over. In the fully actuated system (A), the sweep degree of freedom is at its control limit, indicating that improvements in sweep control effectiveness (*e.g.*, via larger wing chord) would lead to greater pitch control effectiveness. However, the $\Lambda$-$\alpha$ system, case (B), is not at any control limits, indicating that more complex effects are also at play, for example the balance between the lift- and drag-generated pitch-up moment, and the optimisation trade-off that increased sweep represents for these two moments.

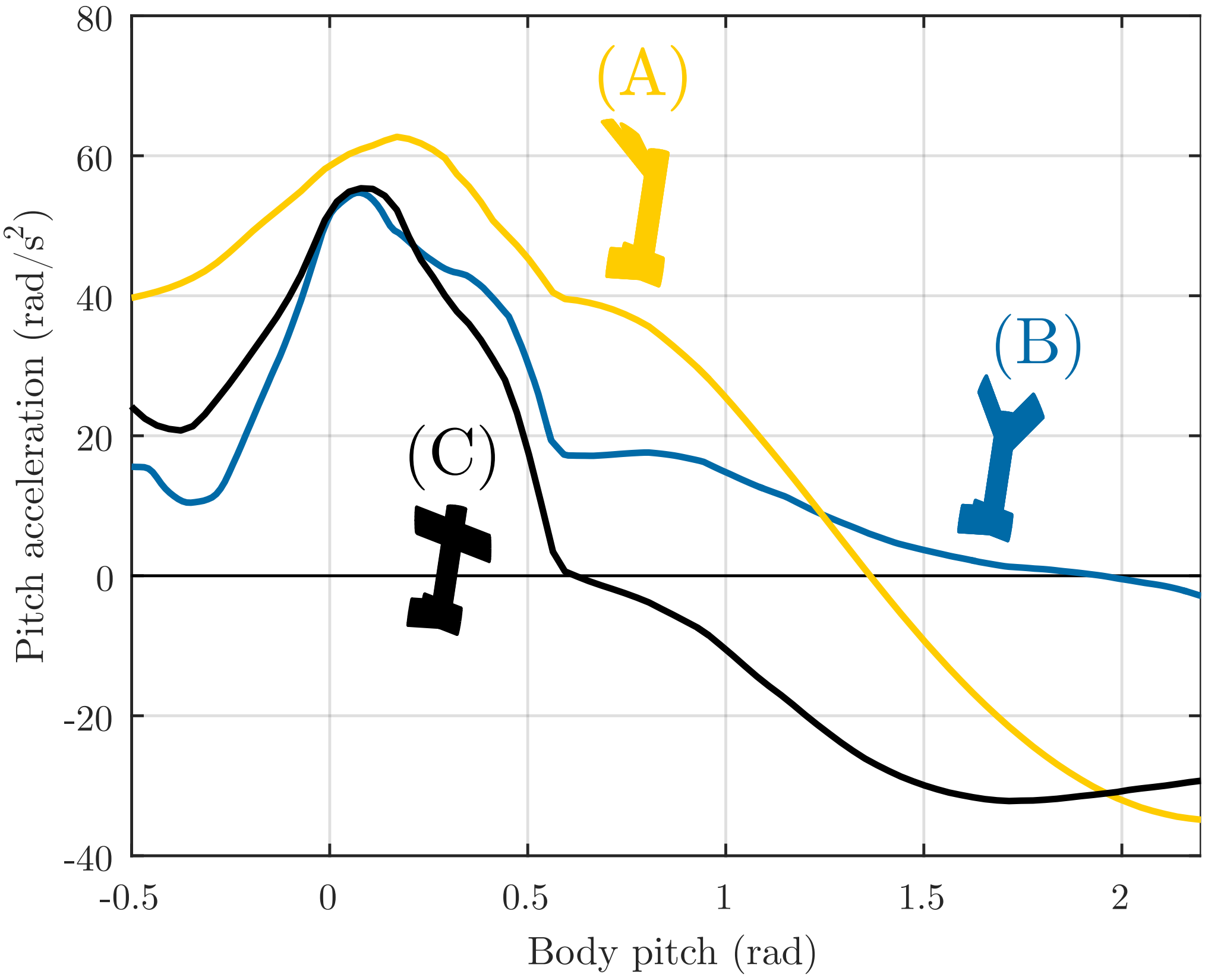

**Figure 6:** Static longitudinal stability profile of several candidate pitch-up configurations: (**A**) with all morphing DOF enabled; (**B**) with only sweep ($\Lambda$) and incidence ($\alpha$) DOF enabled; (**C**) with only the incidence DOF enabled. The key feature of these profiles is the degree to which a positive (upwards) pitch acceleration is maintained at high angles of attack: the longer a positive acceleration is maintained, the greater the maximum attainable angle-of-attack during a RaNPAS manoeuvre.

**Table 4:** Parameters for optimal pitch-up configurations. Values in bold type are located on their respective constraint limits.

| Parameter | **(A)** All DOF | **(B)** $\Lambda$-$\Phi$ | **(C)** $\Phi$ |
|---|---|---|---|
| Dihedral, $\Gamma$ (rad) | 0.730 | 0 | 0 |
| Sweep, $\Lambda$ (rad) | **1.171** | 0.699 | 0 |
| Incidence, $\Phi$ (rad) | 0.247 | 0.181 | 0.171 |
| Elevator deflection, $\beta_e$ (rad) | **−0.870** | **−0.870** | **−0.870** |

As per Fig. 6 and Table 4, forms of wing morphing associated with a high pitch-up rate include positive dihedral, forward wing sweep and mild upwards wing inclination. Upwards wing inclination increases wing lift, but too much reduces the drag-induced pitch-up moment at high angles of attack. Positive dihedral, in combination with forward sweep, induces a drag-based pitch-up drag moment even at lower angles of attack. In addition, forward sweep shifts the aerodynamic centre further forward, increasing its pitching moment about the centre of mass (which is less strongly affected by the sweep motion)—analogous with sweep-based control in birds [69]. The result is that the UAV's stable pitch quasi-trim configuration (the pitch equilibrium $\ddot{\theta} = 0$) is shifted to a very to a high angle of attack—in (B), even to a partially-inverted position. However, while (B) has a quasi-trim configuration at the highest angle of attack, the strength of its attraction is significantly weaker than that of configuration (A), as indicated by the pitch acceleration gradient at the quasi-trim configuration. Configuration (A) is thus likely to allow the cobra manoeuvre to be carried out more rapidly. Note that the use of forward sweep in does have the disadvantage of decreasing the aeroelastic divergence speed of the wings [70], limiting the flight envelope of these forward-swept configurations.

For the pitch-down configuration (3) analogous objective functions are available; though the relevant pitch angles for pitch acceleration minimization are higher (>1 rad), because pitch-down motion must begin from this state. Figure 7 shows optimal configurations to minimise the point pitch acceleration at 1.4 rad (80°) pitch, and Table 5 shows their parameter values. Configurations A-C are the optimal configurations for (A) all degrees of freedom active, (B) sweep and incidence active and (C) only incidence active. Configuration (D) is an example trim configuration, at zero sweep and fuselage angle of attack 0.08 rad (4.6°).

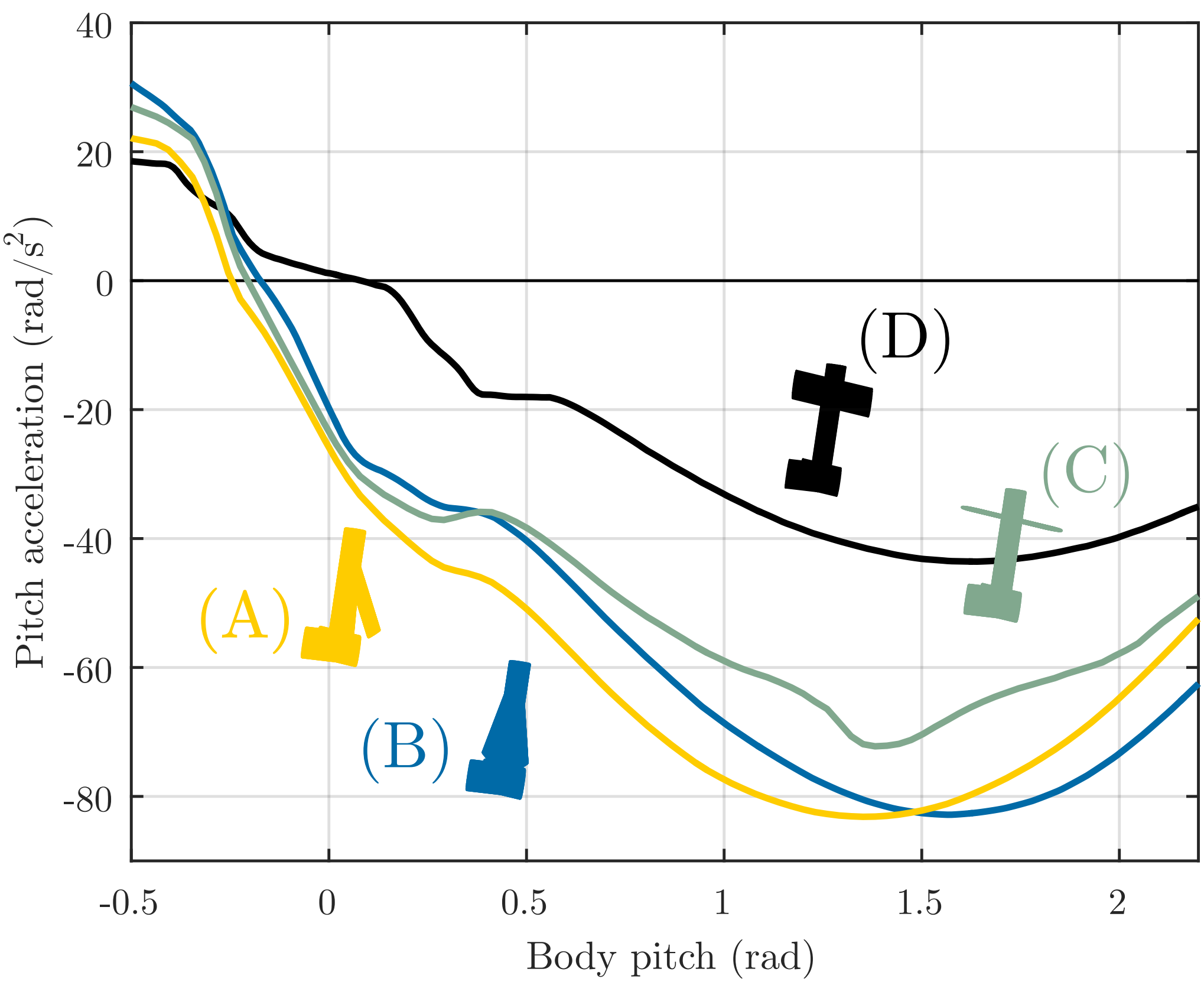

**Figure 7:** Static longitudinal stability profile of several candidate pitch-down configurations: (**A**) with all morphing DOF enabled; (**B**) with only sweep ($\Lambda$) and incidence ($\alpha$) DOF enabled; (**C**) with only the incidence DOF enabled; (**D**) the initial trim configuration. The key feature of these profiles is the strength of the negative (downwards) pitch acceleration at high angles of attack ($> 90°$): the greater the negative pitch acceleration, the more likely that the UAV can recover from a high-angle-of-attack states.

**Table 5:** Parameters for optimal pitch-down configurations. Values in bold type are located on their respective constraint limits.

| Parameter | **(A)** All DOF | **(B)** $\Lambda$-$\Phi$ | **(C)** $\Phi$ | **(D)** Trim |
|---|---|---|---|---|
| Dihedral, $\Gamma$ (rad) | −0.255 | 0 | 0 | 0 |
| Sweep, $\Lambda$ (rad) | **−1.171** | **−1.171** | 0 | 0 |
| Incidence, $\Phi$ (rad) | 0.168 | 0.0493 | −1.654 | 0.014 |
| Elevator deflection, $\beta_e$ (rad) | 0.262 | 0.262 | 0.262 | 0.003 |

Similar aerodynamic effects to those in Figure 6 are observed. Backward sweep moves the aerodynamic centre rearwards, and the presence of anhedral allows the maximum wing surface area to be inclined into the flow, for maximum restoring drag moment. In these cases (configuration A, B), the wing incidence is kept flat to make use of this restoring drag moment; however, when only incidence motion is available (configuration C), inclining the lifting surface into the local airflow to reduce its drag is the better option. The tail then provides all the available restoring moment. Configuration C has the additional benefit of generating significant lift at high angles of attack, reducing the burden on the propulsion system. The trim configuration itself generates moderate pitch-down acceleration; but as per Fig 7 this acceleration can be doubled in the presence of wing morphing.

### 4.2. Flight simulation of a 3DOF-morphing cobra manoeuvre

With candidate pitch-up and pitch-down configurations identified, we simulate several differing forms of cobra manoeuvre. We start with the simplest to define: the UAV begins at an initial trim state; then changed to the 3DOF (all-DOF) pitch-up configuration, as per Fig. 6 and Table 4; and then returns to the original trim state. The only free variables are the timings of the configuration changes, which we define manually. Figure 8 shows the flight simulation results for a simple cobra manoeuvre of this form, including the UAV flight path, its control and orientation history, and its acceleration history compared with the quasistatic states (Fig. 6-7). This simulation is performed under the full GK aerodynamic model. The initial and final near-trim state is the trim state at pitch 0.08 rad and airspeed 30 m/s; with the system initialised at pitch 0.08 rad and airspeed 40 m/s. The time-scales of the morphing motion are 500 ms in the near-trim state, 100 ms transition, 50 ms in the trim-up state and then an immediate return to the near-trim state. This discontinuous control path is then smoothed strongly via a Laplacian smoother, leading to the final control commands of Figure 8.

The manoeuvre is successful: the UAV reaches a nose-up state ($\theta = 1.56$ rad) within half a second of the control onset, losing 19 m/s of airspeed in the process (a reduction of 46%). The UAV then regains airspeed as it transitions into a shallow dive, though not without pitch-down overshoot—it reaches a pitch-down peak of $-0.46$ rad ($-26°$). The manoeuvre is roughly altitude neutral, as altitude gain due to the vertical thrust component at peak pitch offsets altitude loss during dive recovery. Finally, we compare the dynamic pitch acceleration history of the UAV to the quasistatic nonlinear longitudinal stability profiles of the control configurations (Fig. 8D). During the early pitch-up manoeuvre, the quasistatic pitch-up configuration profile predicts the dynamic profile relatively well, but, by the point of peak pitch, the dynamic profiles differ significantly—a difference attributable to dynamic effects and airspeed loss. Despite the difference, the manoeuvre is performed successfully, indicating that heuristics based on longitudinal stability profiles can be a successful strategy for designing supermanoeuvres in biomimetic UAVs

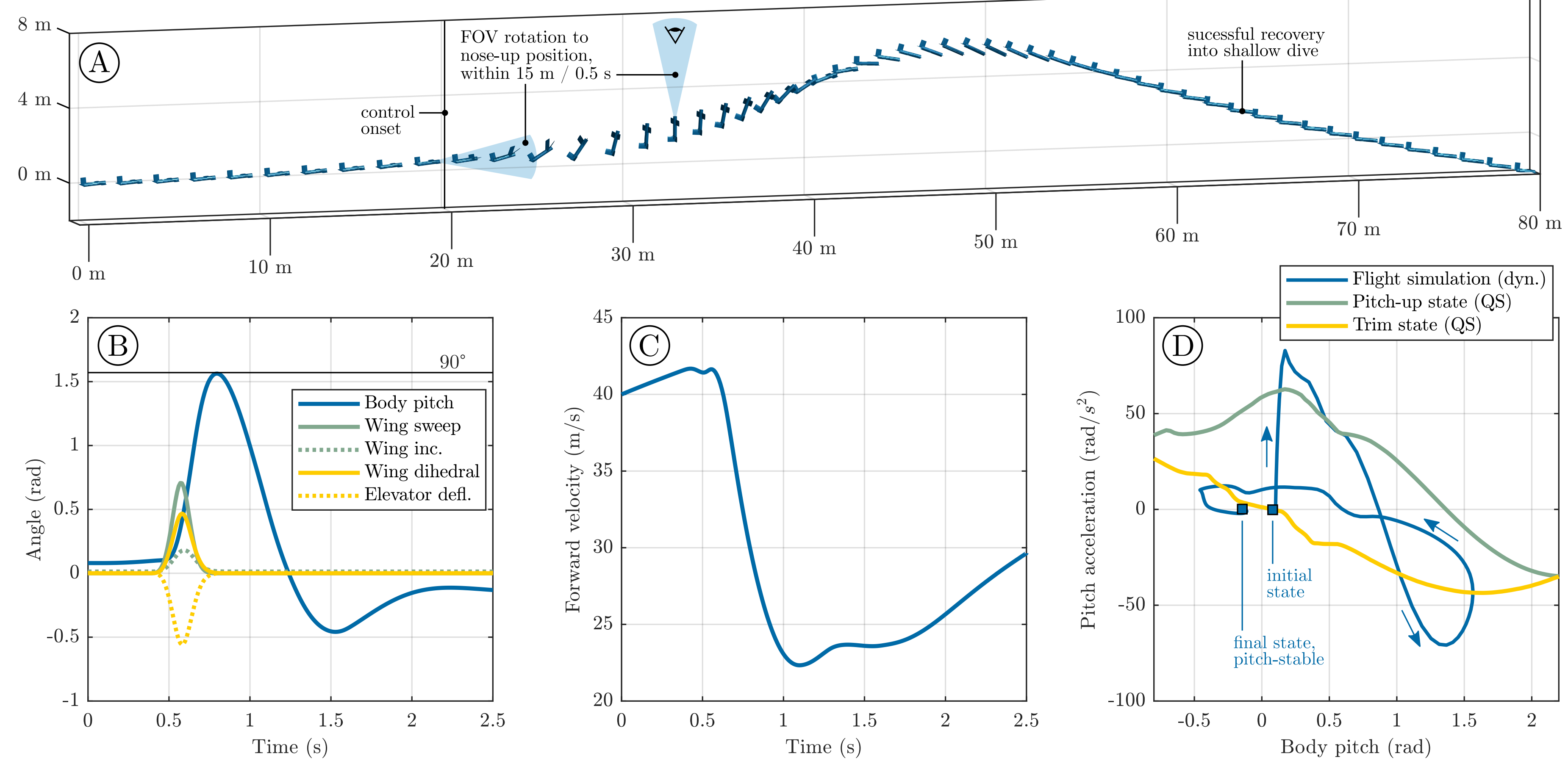

**Figure 8:** Flight simulation results for a simple 3DOF-morphing cobra manoeuvre at T/W = 0.25, under a quasisteady aerodynamic model. (**A**) flight path with UAV rendered every 50 ms ($0 \leq t \leq 2.5$ s); (**B**) control and orientation history; (**C**) forward velocity history; and (**D**) acceleration history compared with the quasistatic acceleration profiles are shown. The UAV configuration sequence is: near-trim → optimal pitchup → near-trim.

### 4.3. Flight simulation of a 2DOF-morphing cobra manoeuvre

The 3DOF-morphing cobra manoeuvre studied in §4.2 is high-performance, but this level of morphing complexity may not be feasible in all UAVs: we are interested in reducing the morphing complexity required to successfully perform a cobra manoeuvre. Considering the candidate configurations studied in §4.1 (Fig. 6-7), we observe that the sweep-incidence ($\Lambda$-$\alpha$) morphing combination can achieve similar levels of pitch-up and pitch-down strength to the full 3DOF combination. In particular, the addition of sweep morphing shifts the quasi-trim point of the pitch-up configuration (Fig. 6) to a point at greater than 90° pitch. Physically, this corresponds to forward motion of the UAV aerodynamic centre, generating strong pitch-up moment. In manoeuvre design terms, this bodes well for the use of sweep-incidence morphing for the generating the required initial pitch-up moment. However, simply chaining together the sweep-incidence candidate states given in §4.1 does not lead to a successful manoeuvre: the pitch-up moment is insufficient to bring the UAV to beyond 90° pitch.

To resolve this issue, we add another manoeuvre component. Previous analysis of this case study morphing-wing UAV [18,46] revealed the existence of a space of morphed trim states across pitch and yaw: a space of states, at different fuselage orientations, that each could represent a steady level flight state. To reduce the pitch-up moment requirement in the main stage of the manoeuvre, we use these morphed trim states to bring the UAV up to its maximum trim state pitch of 0.5 rad (29°) (Fig. 9), over a short duration. From this point we use the candidate pitch-up and pitch-down configurations of §4.1 to generate a cobra manoeuvre. This approach is successful: Fig. 9 shows the resulting set of cobra manoeuvres that can be achieved at differing initial airspeeds. In general, these manoeuvres show better performance than the manoeuvre in Fig. 8: they have a greater peak pitch angle (up to 1.95 rad), minimal pitch-down overshoot during the recovery phase (down to only $-0.096$ rad), and smaller altitude loss. However, they are less rapid, with the peak pitch point occurring within two seconds of the control onset: a consequence of the initial trim-state alteration. This manoeuvre highlights the significance of trim space analysis, as per [18,46], to the study of biomimetic supermanoeuvrability: trim space manoeuvres, or quasistatic NPAS (QNPAS), can form an important component in rapid NPAS (RaNPAS).

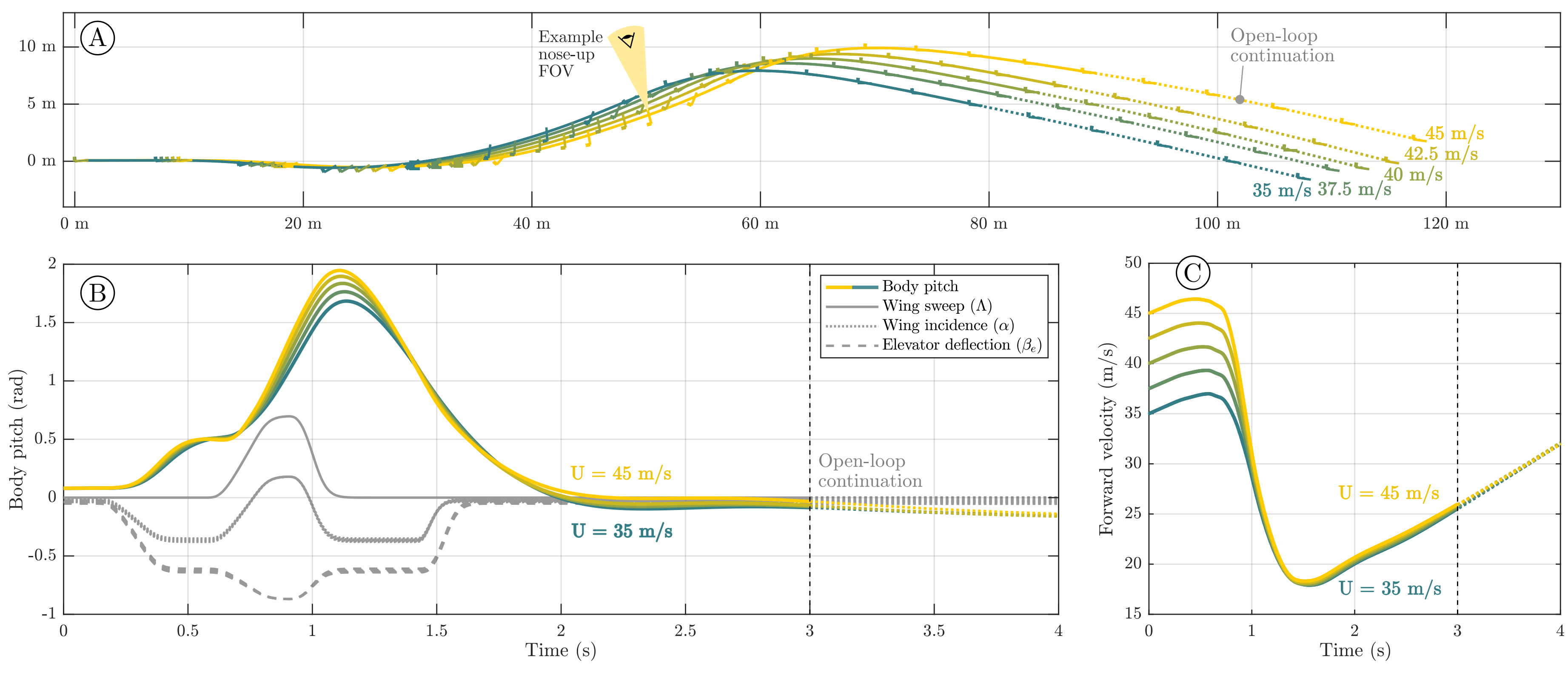

**Figure 9:** Flight simulation results for a 2DOF-morphing cobra manoeuvre at T/W = 0.25, with varying initial airspeed, and using the full GK aerodynamic model. (**A**) flight path with UAV rendered every 200 ms ($0 \leq t \leq 4$ s); (**B**) control and orientation history; (**C**) forward velocity history. The UAV configuration sequence is: near-trim at $\alpha = 0$ rad → trim at $\alpha = 0.4$ rad → optimal pitchup → trim at $\alpha = 0.4$ rad → near-trim at $\alpha = 0$ rad. Beyond $t = 3$ s, the open-loop response of the UAV (a shallow dive) is simulated, as an illustration of the post-manoeuvre recovery process. In reality, beyond $t = 3$ s is the region in which conventional manual or automatic closed-loop flight control would be expected to be reactivated, to purse whatever post-manoeuvre objective is relevant.

# 5. Biomimetic ballistic transition capabilities

## 5.1. Manoeuvre design

To date, biomechanical studies have not identified any forms of animal post-stall manoeuvrability that directly parallel the RaNPAS supermanoeuvres studied in §4. This may be a result of the close association between RaNPAS and equipment or morphology based on field-of-view (cannon, *etc.*) rather than on direct contact (beaks, talons, *etc.*). In animals, the absence of field-of-view weapons would be expected to render true RaNPAS manoeuvres of minimal utility. Nevertheless, some forms of animal manoeuvre show correspondences with more general forms of NPAS capability, though their motivation is not primarily to effect orientation changes. One such manoeuvre is the ballistic transition, observed in a variety of gliding mammals [7–9]. The ballistic transition manoeuvre is a cobra manoeuvre with no pitch-down configuration: the objective is to decrease the airspeed of the animal, or UAV, in preparation for an impact landing on a vertical surface. As such, manoeuvre design for a ballistic transition manoeuvre in our biomimetic UAV can proceed along fundamentally the same lines as §4.1. We utilise a simple modification of the cobra controls: we generate a pitch-up moment via a 2DOF ($\Lambda$-$\alpha$) morphing pitch-up configuration (Table 4, Fig. 6), but rather than transitioning subsequently to a pitch-down configuration, the UAV transitions to a neutral configuration which maintains high pitch angle at decreasing airspeed until the point of impact. One convenient near-neutral configuration is same pitch-up state but with zero incidence and elevator deflection (*i.e.*, only forward sweep). In a manoeuvre of maximum simplicity, this state may be maintained until impact landing—as we will now simulate.

## 5.2. Flight simulation

Figure 10 shows a simulation of a ballistic transition manoeuvre in the biomimetic UAV, under the GK aerodynamic model, and utilising the 2DOF (sweep-incidence, $\Lambda$-$\alpha$) sequence of control configurations studied in Fig. 6. The objective is a low-velocity impact landing on the vertical surface of a building, 45 m away, starting at forward velocity of 60 m/s. The effect of a varying constant thrust value ($0.2 < T/W < 1$) is shown. For all the simulated thrust values, the ballistic transition manoeuvre is successful, for a single set of control timings: the UAV lands in an almost exact vertical position, with both the horizontal and vertical velocity $< 12$ m/s. Even in the worst case, $T/W = 0.2$, the system kinetic energy is reduced by 94% at the point of impact, with near-zero altitude change. The primary effect of T/W is to increase the altitude gain through the manoeuvre, reducing the system kinetic energy via transfer to gravitational potential. Maximal levels of kinetic energy dissipation (up to 97.5%) are thus associated with maximal available thrust (at least, up to T/W = 1). However, the effect of T/W on the aerodynamic energy dissipation is only secondary, as evidenced by the total energy trends: optimal total energy dissipation occurs at T/W = 0.6, but the variation is not large. The use of altitude gain for kinetic energy is useful but (**i**) may not be permissible for impact landing in confined environments, and (**ii**) may be achievable through finer morphing control at lower T/W. We note also that the use of forward sweep, and the corresponding reduction in wing aeroelastic divergence speed, is likely to limit the initial airspeed of the manoeuvre: a more versatile manoeuvre sequence is to use incidence morphing (and, if available, dihedral) for initial airspeed reduction before a sweep morphing phase. Aeroelastic tailoring is also an option to increase the divergence speed, *cf.* [71].

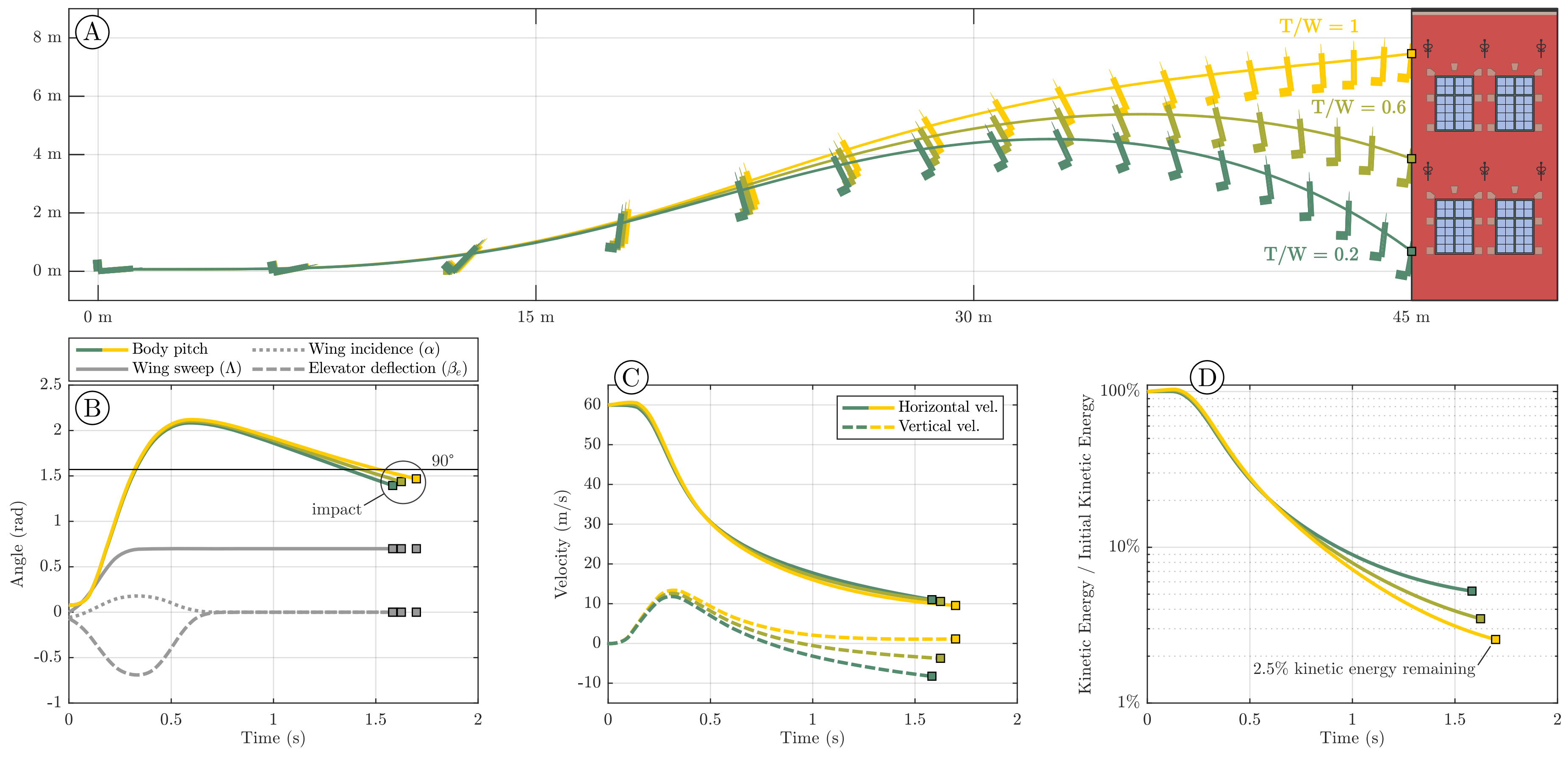

**Figure 10:** Flight simulation results for a 2DOF-morphing ballistic transition manoeuvre with initial velocity 60 m/s, under varying initial thrust (T/W). The UAV configuration sequence is: trim → pitchup → stabilisation state. (**A**) UAV flight paths, overlaid on an illustrative scenario involving landing on a building. (**B**) Body pitch angle histories and control histories, indicating the varying point of impact landing. (**C**) Horizontal and vertical velocity histories. (**D**) Relative kinetic energy history, indicating that in the best case (T/W = 1), the impact landing occurs with only 2.5% of the UAV's initial kinetic energy.

## 6. Effects of aerodynamic model fidelity

### 6.1. Model fidelity effects in a cobra manoeuvre context

The simulations in §§4-5 were performed with the extended GK dynamic stall model (§3), which captures dynamic stall effects on all UAV lifting surfaces. We ask two questions regarding the role of dynamic stall effects in these manoeuvres. *Firstly*: how significant are dynamic stall effects in these manoeuvres? And *secondly*: where are these manoeuvres located within the windows of quasisteady and GK model validity outlined in §3.4? Considering first the 3DOF cobra manoeuvre of §4.2, Fig. 11 illustrates flight simulation results (pitch angle and flight path) for three different aerodynamic models: (**i**) the quasisteady model based on original source data; (**ii**) the GK-reconstructed quasisteady model, with $p = p_0$ (§§3.2-3.3); and (**iii**) the transient GK model with full dynamic stall effects. Figure 9 also illustrates wing- and stabiliser-tip lift coefficient histories, and includes an assessment of the full transient GK simulation in terms of reduced frequency and reduced pitch rate, with the model validity thresholds noted.

As can be seen in Fig. 11, the cobra manoeuvre in this biomimetic UAV is remarkably resilient to dynamic stall: while dynamic stall causes significant changes in the manoeuvre lift coefficient peaks, these changes do not fundamentally alter the manoeuvre. A pair of explanations for this resilience are available. (**i**) The cobra manoeuvre may show intrinsic high levels of stability in simulation, arising from the planar nature of the manoeuvre—eliminating the possibility of roll drift. This may explain why the large but short-timescale lift peaks arising from delayed separation do not significantly perturb the manoeuvre. And (**ii**) the symmetric nature of the hysteresis loop (with delayed stall and delayed reattachment) may serve to self-cancel in a sufficiently stable system. The planar assumption obviously neglects more complex transient effects arising from asymmetric forebody separation [72,73]; if uncontrolled, these could initiate a GK-based destabilisation. The uncertain nature of these effects implies that, in a real UAV, they may be more significant, and may require a control response.

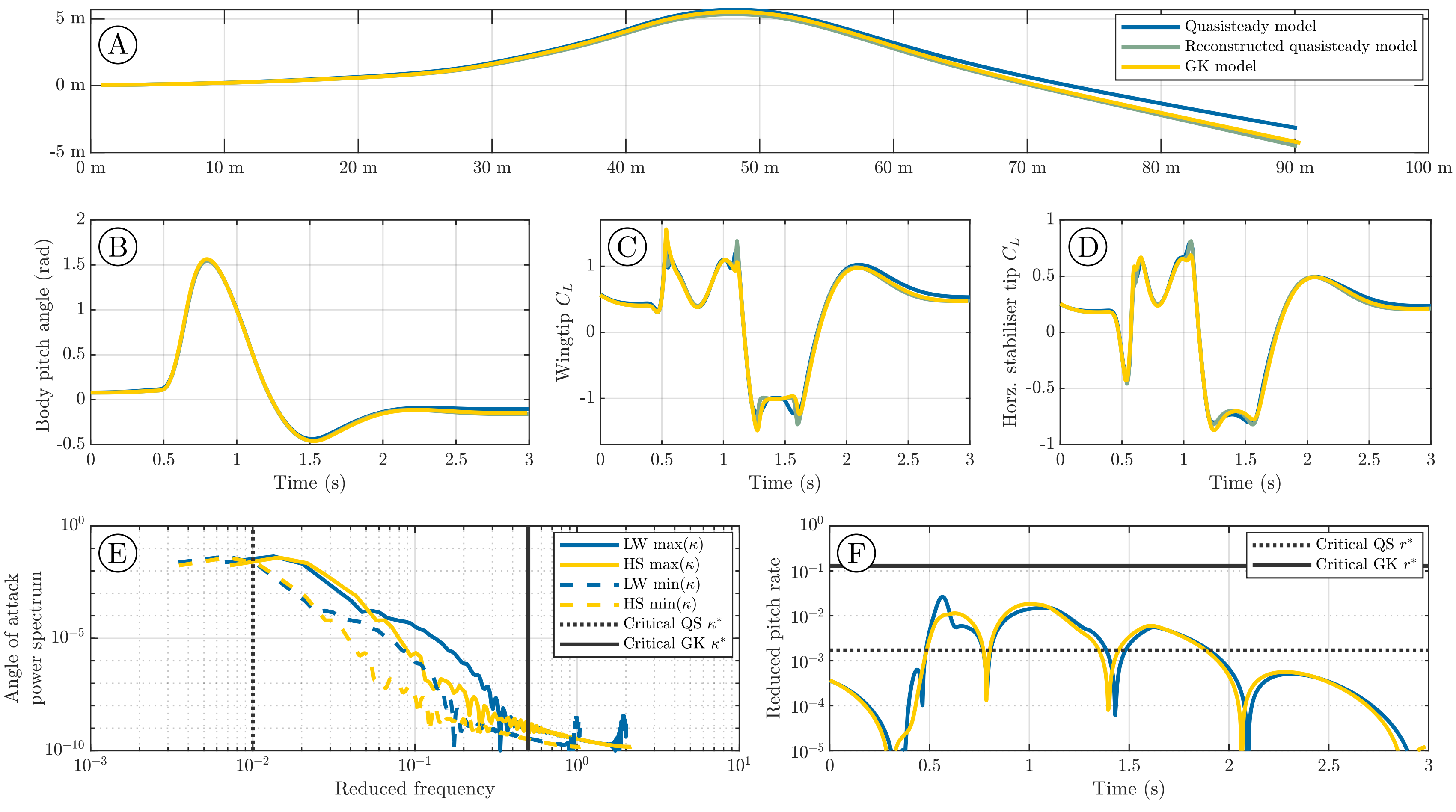

**Figure 11:** Validation flight simulation results for a simple 3DOF-morphing cobra manoeuvre at T/W = 0.25: simulations with the quasisteady (QS) aerodynamic model; with the GK-reconstructed quasisteady aerodynamic model; and with the full GK model (lift, drag, moment). (**A**) Flight path; (**B**) Orientation history; (**C**) Wingtip lift coefficient; (**D**) Horizontal stabiliser tip lift coefficient; (**E**) Angle of attack power spectrum, indicating the approximate limits of QS and GK model validity. (**F**) Reduced pitch rate profile, indicating the approximate limits of QS and GK model validity. As can be seen, the manoeuvre lies within the limits of GK model validity; and, despite lying partly outside the limits of QS model validity, is well-approximated in simulation.

### 6.2. Model fidelity effects in a ballistic transition manoeuvre context

We perform the same analysis on a ballistic transition manoeuvre, as per §5. Taking the highest-performance manoeuvre at T/W = 1, we perform the same three simulations: (**i**) with the quasisteady aerodynamic model based on original source data; (**ii**) with the GK-reconstructed quasisteady model, with $p = p_0$ (§§3.2-3.3); and (**iii**) with the full transient GK model. Figure 12 shows the results of these simulations, including an assessment of the full transient GK simulation in terms of reduced frequency and reduced pitch rate, with the thresholds for GK and quasisteady model validity noted. The results are notably different to the cobra manoeuvre results of §6.1. The same relatively brief differences in lift coefficient peak (due to lift hysteresis) are present, but in the ballistic transition manoeuvre, they have a significant effect. Lift hysteresis effects alter the aerodynamic behaviour of both the wings and horizontal stabiliser during the pitch-up segment of the manoeuvre ($0 \leq t \leq 0.5$); an alteration which then propagates to significant changes in the UAV flight path, and its kinetic energy at the point of impact landing.

These changes are particularly interesting because the manoeuvre unsteadiness, as measured by the reduced pitch rate, is not extreme—only just straying into the region in which the quasisteady model is expected to be invalid ($r \geq 0.0017$). In practical terms, unlike the cobra manoeuvre studied in §6.1, this ballistic transition manoeuvre is not resilient to aerodynamic uncertainty (*vis-à-vis* lift hysteresis), perhaps due to the absence of any pitch-down manoeuvre segment which might serve as a self-cancellation mechanism. This highlights the importance of considering dynamic stall effects when designing and controlling ballistic transition manoeuvres in biomimetic and other UAVs. Note also that further out-of-plane effects, such as asymmetric forebody separation, may also be at work, and may require a control response. Asymmetric forebody separation in biomimetic UAVs is an interesting topic for future research: existing studies, focusing on crewed combat aircraft, typically consider delta wing geometries [72–75], rather than the forward-swept wing configuration used in this biomimetic UAV. The extent to which the results of these analyses apply in the forward-swept case is currently uncertain.

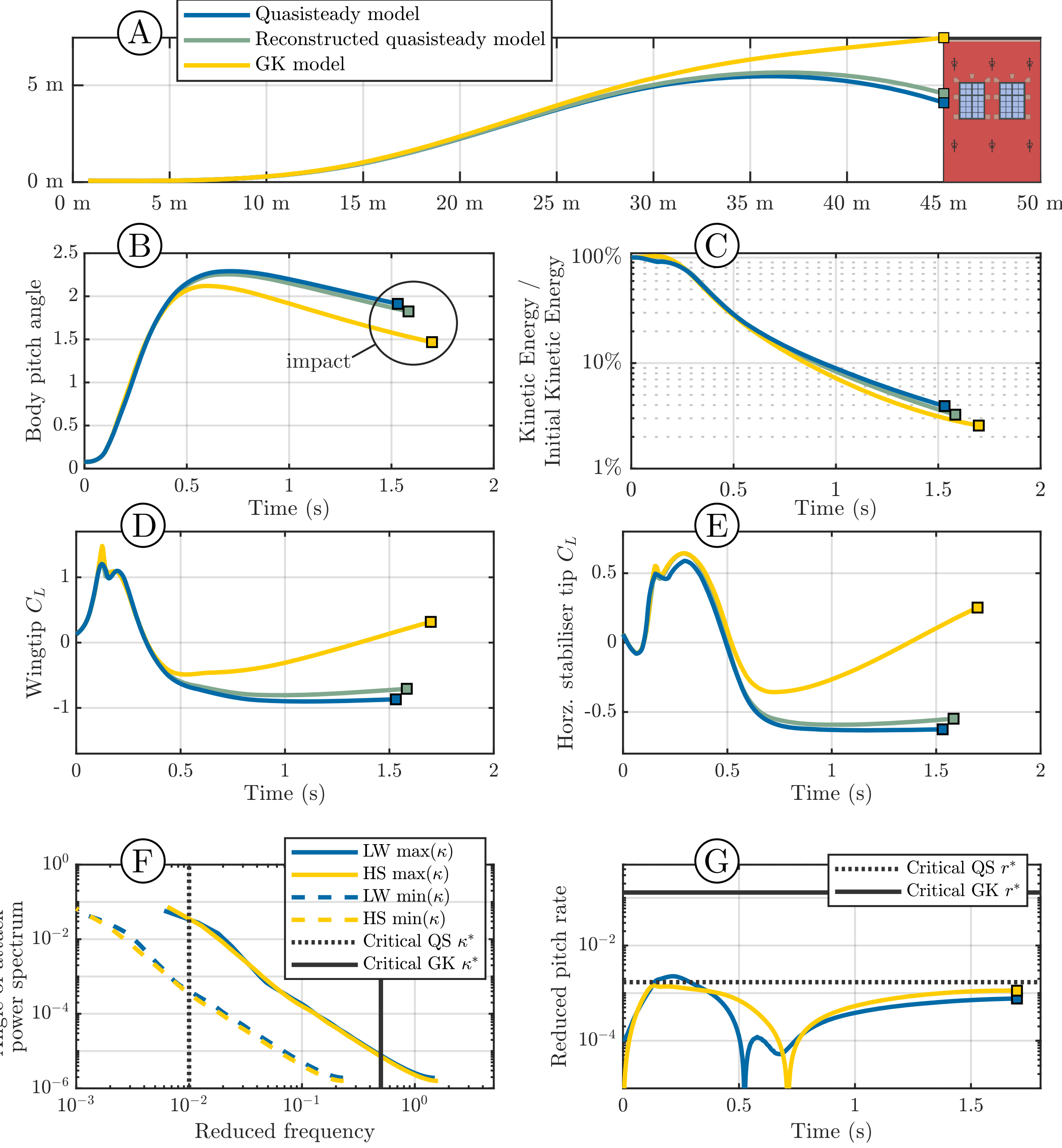

**Figure 12:** Validation flight simulation results for a ballistic transition manoeuvre at T/W = 1: simulations with the quasisteady (QS) aerodynamic model; with the GK-reconstructed quasisteady aerodynamic model; and with the full GK model (lift, drag, moment). (**A**) Flight path; (**B**) Orientation history; (**C**) Relative kinetic energy history; (**D**) Wingtip lift coefficient; (**E**) Horizontal stabiliser tip lift coefficient; (**F**) Angle of attack power spectrum, indicating the approximate limits of QS and GK model validity. (**G**) Reduced pitch rate profile, indicating the approximate limits of QS and GK model validity. As can be seen, the manoeuvre lies within the limits of GK model validity; and, despite lying partly outside the limits of QS model validity, is well-approximated in simulation.

Finally, we observe that, in the ballistic transition manoeuvre, the reduced frequency ($\kappa$) is not a reliable metric of lifting surface aerodynamic transience: the UAV airspeed changes significantly, leading to significant uncertainty in the value of the appropriate local airspeed ($U$) used in the reduced frequency calculation ($\kappa = b\Omega/U$). Figure 12 shows coarse estimates based on maximum and minimum airspeed across the manoeuvre. As a result of the time-frequency uncertainty principle (Gabor limit) [76], it is impossible to locate spectral components of a signal ($\Omega$) precisely in time ($t$, and therefore $U$), and thus impossible to provide exact manoeuvre time-histories for reduced frequency $\kappa$. However, improved estimates may be available via the use of short-time Fourier transforms [77].

## 7. Discussion and conclusion

In this work, we have demonstrated a biomimetic mechanism for enabling both classical and bio-inspired pitch-axis supermanoeuvrability in a UAV. Pitch-axis RaNPAS, in the form of the cobra manoeuvre, is demonstrated in multiple manoeuvre variants and across multiple aerodynamic models. This RaNPAS capability is achievable without thrust vectoring: it is instead achieved via biomimetic wing morphing. While there are several limitations regarding closed-loop control, and out-of-plane aerodynamic effects, this novel demonstration that classical supermanoeuvrability is possible via biomimetic wing morphing is an advance in our understanding of aerial manoeuvrability. It provides an impetus for considering the role of supermanoeuvrability in autonomous dogfighting, with the potential to improve dogfighting performance both by introducing novel manoeuvres and by confounding adversary manoeuvre predictive tracking [78]. This demonstration also raises several further lines of research, and open questions. In particular, we raise the question of the relative advantages of thrust-vectoring supermanoeuvrability *vis-à-vis* morphing-wing supermanoeuvrability. Are there manoeuvres which are possible via one of these mechanisms, but not the other? We can identify several contexts in which morphing-wing supermanoeuvrability might be preferable: this mechanism does not rely on thrust to effect orientation changes, and thus requires relatively low UAV thrust-to-weight ratios—commensurate with lower-performance UAVs currently in use. In addition, variants of these manoeuvres are likely to be possible without propulsion, at a glide state. Morphing-wing supermanoeuvrability is also directly relevant to the control of flapping-wing micro-air-vehicles (FW-MAVs). The guidance method demonstrated in this paper, based on control of the UAV longitudinal stability profile, could enable exploration of the space of supermaneuvers available to FW-MAVs.

In addition, in this work we also explored the connection between classical supermanoeuvrability and animal flight manoeuvrability: we demonstrated how an extreme animal flight manoeuvre—ballistic transition—is available in the same biomimetic UAV, and can be generated via the same manoeuvre design strategy. Indeed, in this biomimetic UAV, the ballistic transition is found to be related to the classical cobra: both can be achieved via a common pool of morphing configurations. This not only highlights the commonality between UAV supermanoeuvrability and animal manoeuvrability, but paves the way for translating animal manoeuvrability to UAVs: for instance, UAVs capable of rapid landing on vertical or inverted surfaces. In this work, we focused on pitch-axis manoeuvrability, but further study of biological flight manoeuvrability along other flight axes—including biological stall turns [1–3], and zero-airspeed rolling manoeuvres [6]—could lead to further novel forms of UAV manoeuvrability. The intersection between classical supermanoeuvrability, biological flight manoeuvrability, and morphing-wing aircraft is a cross-disciplinary topic with the potential to significantly advance our understanding of aerial manoeuvrability and autonomous dogfighting; and to lead to new designs of UAV.

**Acknowledgements.** AP was supported by the Cambridge Commonwealth Trust, Commonwealth Prince of Wales Scholarship.

**Competing interests.** The authors declare that they have no known competing financial interests or personal relationships that could have appeared to influence the work reported in this paper.